\documentclass[10pt,final,doublecolumn]{IEEEtran}
\usepackage{amsmath}
\usepackage{amssymb}
\usepackage{amsfonts}
\usepackage{latexsym}
\usepackage{graphicx}
\usepackage{bbding}
\usepackage{enumerate}
\usepackage{subeqnarray}
\usepackage{multicol}
\usepackage{color}
\usepackage{setspace}
\usepackage{mathrsfs}
\usepackage{array}
\usepackage{algorithm,algorithmic}
\usepackage{colortbl}
\usepackage{bm}
\usepackage{setspace}
\usepackage{url}
\IEEEoverridecommandlockouts
\allowdisplaybreaks[3]

\begin{document}
\title{Integration of Navigation and Remote Sensing in LEO Satellite Constellations}
\author{Qi Wang, Xiaoming Chen, Qiao Qi, Zhaolin Wang, and Yuanwei Liu
\thanks{Qi Wang and Xiaoming Chen are with the College of Information Science and Electronic Engineering, Zhejiang University, Hangzhou 310027, China (e-mails: wang-qi@zju.edu.cn; chen\underline{~}xiaoming@zju.edu.cn). Qiao Qi is with the School of Information Science and Technology, Hangzhou Normal University, Hangzhou 311121, China (e-mail: qiqiao@hznu.edu.cn). Zhaolin Wang is with the School of Electronic Engineering and Computer Science, Queen Mary University of London, E1 4NS London, U.K. (e-mail: zhaolin.wang@qmul.ac.uk). Yuanwei Liu is with the Department of Electrical and Electronic Engineering, the University of Hong Kong, Hong Kong, China (e-mail: yuanwei@hku.hk).
}}\maketitle

\begin{abstract}
Low earth orbit (LEO) satellite constellations are becoming a cornerstone of next-generation satellite networks, enabling worldwide high-precision navigation and high-quality remote sensing. This paper proposes a novel dual-function LEO satellite constellation frame structure that effectively integrating navigation and remote sensing. Then, the Cramer-Rao bound (CRB)-based positioning, velocity measurement, and timing (PVT) error and the signal-to-ambiguity-interference-noise ratio (SAINR) are derived as performance metrics for navigation and remote sensing, respectively. Based on it, a joint beamforming design is proposed by minimizing the average weighted PVT error for navigation user equipments (UEs) while ensuring SAINR requirement for remote sensing. Simulation results validate the proposed multi-satellite cooperative beamforming design, demonstrating its effectiveness as an integrated solution for next-generation multi-function LEO satellite constellations.
\end{abstract}


\section{Introduction}
In recent years, the development from ground-based services to space-based systems has undergone significant technological evolution and breakthroughs as the global demand for information services continues to grow. Early ground-based services relied on navigation stations and observation sites, but their limited geographic coverage and performance bottlenecks made it increasingly challenging to meet the complex and diverse needs of a globalized world \cite{ground-based services}. The emergence of satellite technology offers a transformative solution, enabling remote navigation, sensing, and earth observation. However, traditional satellite systems gradually revealed their limitations, including restricted coverage, slower response times, and constrained service capabilities. In this context, low earth orbit (LEO) satellites have attracted widespread attention. Orbiting at altitudes between 200 and 2,000 kilometers, LEO satellites deliver low-latency, high-resolution, and rapid-response services, effectively addressing the limitations of traditional satellite systems \cite{LEO satellites}. Compared to geosynchronous orbit (GEO) and medium earth orbit (MEO) satellites, LEO satellites feature reduced propagation delays, lower path loss, and improved spatiotemporal resolution, making them ideal for precise navigation and remote sensing \cite{compare GEO MEO}.

Typically, the deployment of a single satellite requires the selection of a suitable altitude in order to make a trade-offs between coverage area and signal latency, whereas LEO satellite constellations overcome these limitations through the cooperative operation of multiple LEO satellites in precisely designed orbits, achieving both expanded coverage and reduced latency. This breakthrough has driven a global surge in LEO satellite network development over the past decade, with several large-scale LEO satellite constellations like OneWeb and Starlink successfully deployed worldwide \cite{project1}. With the advent of the LEO satellite constellation era, the collaborative operation of multiple satellites enables seamless global coverage, high-precision navigation, and real-time earth observation, revolutionizing space-based service models \cite{LEO Satellite Constellation}. Additionally, advancements in technology and declining satellite launch costs have further accelerated this trend. Innovations such as the manufacturing of small, cost-effective satellites, reusable rocket technology, and enhanced on-orbit management capabilities have strongly supported the development of global LEO satellite constellations \cite{satellite cost}. In the context of sixth generation (6G) wireless networks, LEO satellite constellations have emerged as the cornerstone of global navigation and remote sensing services. Looking ahead, the number of LEO satellites is expected to grow towards the capacity limits of orbital space, potentially reaching millions \cite{6G satellite1}. This expansion will establish a transformative platform for emerging applications in fields such as industry, transport and environmental management, further accelerating the realization of a fully interconnected and intelligent global society.

For a long time, navigation and remote sensing, as two key application areas of satellite technology, have been the focus of extensive research in both academia and industry. In general, different types of satellites provide services to user equipments (UEs) according to their specific purposes. Navigation satellites focus on positioning, velocity measurement, and timing (PVT), with their development evolving from early GEO navigation satellites to the comprehensive deployment of global navigation satellite systems (GNSS) centered around MEO \cite{GEO_MEO}. This progression has continuously advanced the precision of positioning technology, the stability of velocity measurement, and the reliability of timing services. Pseudo-range measurement, a fundamental technology in satellite navigation, determines the distance between the UE and the satellite by measuring the signal transmission time and incorporating satellite orbital data, forming the basis for high-precision navigation \cite{Pseudo-range measurement}. Further, maximum likelihood estimation (MLE) techniques have emerged as robust solutions for mitigating multi-path effects and combating interference, significantly improving navigation accuracy in challenging environments \cite{GNSS MLE}. Furthermore, the incorporation of LEO satellites has served as a complementary enhancement to traditional navigation systems in highly dynamic environments. Their rapid response and wide-area coverage capabilities have significantly improved the performance of PVT services. For example, the authors in \cite{LEO navigation1} explored the potential of combining LEO satellites with PVT-navigation systems and provided a comprehensive discussion of LEO satellite navigation system design steps, technical challenges, physical layer parameters, optimization tools, propagation models and application scenarios. The authors in \cite{LEO navigation2} optimized the beam direction and scheduling strategy of multi-beam LEO satellites to enhance the user's navigation accuracy while suppressing inter-beam interference. In \cite{LEO navigation3}, the authors proposed a quadruple-coverage constellation design method for LEO navigation satellites by analysing the Walker constellation using the street-of-coverage (SOC) method in order to efficiently achieve global navigation by LEO satellites.

Meanwhile, remote sensing satellites play a crucial role in data acquisition and earth observation, with their technological capabilities evolving from early static imaging to today's dynamic high-resolution monitoring, achieving remarkable progress. Modern remote sensing technologies are widely applied in environmental monitoring, precision agriculture, and disaster assessment, providing essential data support for global change research, resource management, and urban planning. For example, multi-spectral and hyper-spectral remote sensing technologies enable detailed characterization of surface features, monitoring vegetation health, crop growth, and soil moisture \cite{multi-spectral}. Additionally, synthetic aperture radar (SAR) imaging, as an active remote sensing method, uses microwave signals to acquire high-resolution surface information under all-weather and all-time conditions \cite{SAR}. Moreover, LEO satellite remote sensing, known for its significant advantages in dynamic and high-frequency monitoring, has been widely studied. For instance, in \cite{LEO remote sensing1}, the authors proposed a resilient network architecture for LEO remote sensing satellite networks, improving reliability and efficiency through dynamic routing, data caching, and hop-by-hop transmission. The authors in \cite{LEO remote sensing2} presented a global remote sensing framework using LEO satellite constellations integrated with in-orbit cloud computing and AI for real-time earth monitoring and data analysis.

Traditionally, single-function navigation or remote sensing satellites often require independent hardware design, orbital deployment, and ground processing systems. This not only substantially increases the construction and operational costs of satellite systems but also hinders the efficient utilization of resources and the coordination of functionalities. However, with continuous advancements in satellite technology, particularly the significant improvement in payload processing capabilities, the deep integration of navigation and remote sensing technologies has gradually become possible \cite{payload}. This integration, achieved through unified hardware platform design and the collaborative development of multi-functional payloads, enables a single satellite to simultaneously perform navigation and remote sensing tasks, significantly reducing launch frequency and system costs. Additionally, the sharing and optimized allocation of spectrum resources further enhance system operational efficiency. The integration of navigation and remote sensing not only leverages navigation technologies to provide high-precision geo-referencing for remote sensing data, ensuring spatiotemporal consistency in observations, but also utilizes remote sensing technologies to enhance the environmental awareness of navigation systems, providing critical support for precise positioning and path planning in complex scenarios. This integrated technology exhibits enormous potential in fields such as climate change mitigation, global resource management, and sustainable urban development. Specifically, for disaster monitoring, the integrated system can provide rapid and accurate geospatial data for emergency response and damage assessment. In autonomous navigation, it can offer highly precise positioning and environmental awareness, crucial for self-driving vehicles and aerial drones operating beyond line-of-sight. Furthermore, for urban planning, real-time high-resolution remote sensing data combined with precise navigation can enable dynamic infrastructure monitoring and optimized resource allocation, while driving the evolution of satellite systems toward greater intelligence and multi-functionality \cite{navigation and remote sensing1}. In recent years, academic research on integrated navigation and remote sensing satellites has been increasing. For example, the authors in \cite{navigation and remote sensing2} summarised the theories and applications of navigation and remote sensing fusion, proposed a unified representation method, and suggested promoting deep integration through system design to overcome technical bottlenecks. In \cite{navigation and remote sensing3}, the authors explored a GNSS-SAR design algorithm that integrates SAR and navigation systems through joint optimization and satellite trajectory analysis, enabling efficient ground deformation monitoring and supporting disaster response and dynamic target tracking. Meanwhile, integrated sensing and communication (ISAC) and dual-function radar-communication (DFRC) systems, extensively studied in terrestrial networks, focus primarily on integrating radar and communication functions to enhance spectrum and hardware utilization \cite{ISAC1, ISAC2}. However, terrestrial-based ISAC solutions face inherent limitations such as restricted spatial coverage and relatively static deployments. In satellite scenarios, although GEO/MEO satellites support wider coverage, their ISAC implementations suffer from significant propagation delays, reduced spatial-temporal resolution, and limited flexibility due to fixed orbital configurations. In contrast, LEO satellite constellations uniquely overcome these limitations by leveraging dense satellite deployment, cooperative operation, and proximity to Earth, enabling enhanced flexibility, ultra-low latency, and high spatial-temporal resolution.

Nevertheless, most existing studies on integrated navigation and remote sensing primarily focus on single-satellite solutions, which are inherently constrained by limited spatial coverage, infrequent observation intervals, and rigid hardware configurations. Although these approaches demonstrate the feasibility of performing dual functions on a single satellite platform, their performance is fundamentally restricted by the trade-off between coverage area and temporal resolution, as well as by inefficiencies in spectrum and hardware utilization. In contrast, LEO satellite constellations offer unique advantages for integrated navigation and remote sensing. Through the dense deployment and cooperative operation of multiple satellites, LEO satellite constellations can achieve seamless global coverage, ultra-low latency, and high-frequency revisit capabilities, which are unattainable for single-satellite systems. The collaborative multi-satellite architecture enables flexible resource allocation, dynamic beamforming, and effective interference mitigation between navigation and sensing functions. Furthermore, the joint use of hardware and spectrum resources across multiple satellites enhances overall system efficiency and reduces deployment costs. It is also worth noting that practical wireless systems are inevitably affected by hardware impairments such as amplifier nonlinearity and phase noise, which may degrade system performance. Recent studies \cite{hardware impairments1}-\cite{hardware impairments3} explored these effects in reconfigurable surface and satellite-based architectures, providing useful insights for the design of robust dual-function systems. In this context, we attempt to establish a universal framework for integrating navigation and remote sensing functionalities within LEO satellite constellations. By analyzing the key performance metrics and impact factors required of the dual function, we further explore the joint design strategy of beamforming for navigation and remote sensing. The design aims to deliver high-precision navigation services to ground UEs while providing high-quality remote sensing observations for a specific area, thereby driving the comprehensive advancement of LEO satellite constellation technologies. The main contributions of the paper are summarized as follows.

\begin{enumerate}
\item
We present a unified dual-function LEO satellite constellation framework that provides navigation services to UEs while performing remote sensing for a specific ground area by utilizing shared hardware and spectrum resources.

\item
We derive the Cramer-Rao bound (CRB)-based PVT error of a hybrid navigation algorithm combining pseudo-range measurements and MLE as the performance metric for navigation, while defining the closed-form expression of signal-to-ambiguity-interference-noise ratio (SAINR) as the performance metric for remote sensing.

\item
We propose a multi-satellite cooperative beamforming design for integrated navigation and remote sensing in LEO satellite constellations, optimizing both navigation and remote sensing beamforming to enhance navigation accuracy while ensuring remote sensing signal quality.

\end{enumerate}

The subsequent sections are structured as follows: In Section II, we present the system model for LEO satellite constellation integrating navigation and remote sensing. Section III discusses a multi-satellite cooperative beamforming design for integrated navigation and location sensing in LEO satellite constellations. Section IV presents simulation results to demonstrate the effectiveness of the proposed algorithms. Finally, Section V provides the concluding remarks of this paper.

\emph{Notations}: Scalars, vectors, and matrices are represented by ordinary letters, bold lowercase letters, and bold uppercase letters, respectively. The symbols $(\cdot)^T$, $(\cdot)^H$, $(\cdot)^{-1}$, and $(\cdot)^{\dagger}$ denote the transpose, conjugate transpose, inverse, and pseudoinverse of a matrix, respectively. The rank and trace of a matrix are indicated by $\text{Rank}(\cdot)$ and $\text{tr}(\cdot)$, while $\|\cdot\|$ refers to the 2-norm of a vector. $[\mathbf{x}]_i$ is the $i$-th element of the vector $\mathbf{x}$, $[\mathbf{X}]_{i,j}$ is the element in the $i$-th row and $j$-th column of matrix $\mathbf{X}$, $\text{Diag}(\cdot)$ generates a diagonal matrix, $\mathbf{X}\succeq\mathbf{0}$ signifies that matrix $\mathbf{X}$ is positive semi-definite. Horizontal and vertical concatenations of matrices are written as $\left[ {{\bf{X}},{\bf{Y}}} \right]$ and $\left[ {{\bf{X}};{\bf{Y}}} \right]$, respectively. The sets of complex and real matrices with dimensions $\emph{a}$ $\times$ $\emph{b}$ are denoted by ${{\mathbb{C}}^{a\times b}}$ and ${{\mathbb{R}}^{a\times b}}$. The Hadamard product and the Kronecker product are indicated by $\odot$ and $\otimes$, respectively. The cross product between two 3-dimensional vectors is denoted by $\times$. Lastly, $J_1(\cdot)$ and $J_3(\cdot)$ denote the first-order and third-order Bessel functions.

\section{System Model}
\begin{figure}[!t] \centering
\includegraphics [width=0.43\textwidth] {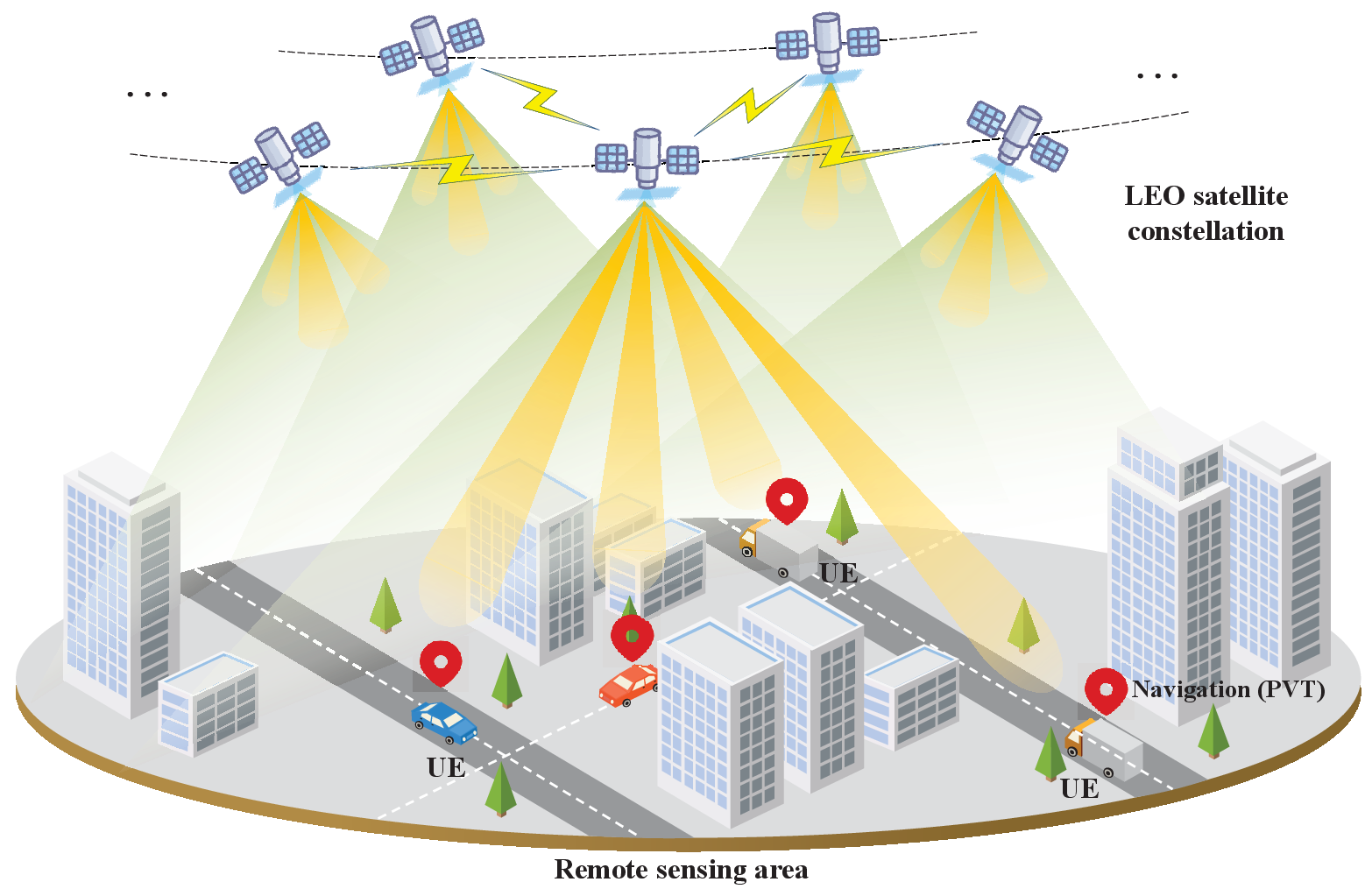}
\caption {System model for LEO satellite constellation integrating navigation and remote sensing.}
\label{system model}
\end{figure}
Consider a LEO satellite constellation designed for continuous global coverage, with the Walker Delta configuration\footnote{Walker Delta is a satellite constellation configuration, where satellites are distributed across multiple orbital planes with equal inclination, uniform spacing, and specific phasing to achieve continuous global coverage. In this setup, the constellation consists of $N^*$ satellites evenly distributed across $P^*$ circular orbital planes centered on earth. All satellites share the same orbital altitude $h^*$ and inclination angle $i^*$, with a phase factor $F^*$ defining the relative phase shift between satellites in adjacent orbital planes. This configuration offers exceptional benefits for both regional and global coverage, making it widely applied in satellite communications, remote sensing, and global navigation systems \cite{Walker}.}. As illustrated in Fig. \ref{system model}, each satellite in the LEO satellite constellation is equipped with a geocentric pointing uniform antenna array (UPA) consisting of $N$ antennas to simultaneously provide navigation and remote sensing services to the ground. Specifically, based on the distribution of navigation user equipments (UEs) and remote sensing area, a joint service group is formed by $K$ dual-function LEO satellites within the LEO satellite constellation. The service group consists of a central satellite and $K-1$ auxiliary satellites, located in close proximity within the same orbit or in neighboring orbits. These satellites transmit dual-function navigation and remote sensing signals over the same spectrum, providing navigation service for $M$ single-antenna UEs while collaboratively conducting remote sensing for a specified ground area. For navigation, all UEs receive and then decode the dual-function signals transmitted through the satellite-terrestrial channel to realize precise PVT. For remote sensing, the central satellite with full-duplex UPA receives and processes the signals reflected by the specified ground area to support applications such as environmental monitoring and urban planning. For ease of reference, the key notations used in this paper are summarized in Table I.

\begin{table}[ht]
\small
\centering
\caption{Summary of Key Notations}
\begin{tabular}{l p{6.4cm}}
\hline
\textbf{Notation} & \textbf{Description} \\
\hline
$K$ & Number of LEO satellites \\
$M$ & Number of UEs \\
$N$ & Number of UPA antennas \\
${{\bf{p}}_m}$, ${{\bm{\gamma }}_m}$ & Position and velocity of the $m$-th UE \\
${{\bf{q}}_k}$, ${{\bm{\eta }}_k}$ & Position and velocity of the $k$-th satellite \\
$\mathbf{v}_{k,m}$ & Navigation beamforming vector at the $k$-th satellite for the m-th UE\\
$\mathbf{w}_k$ & Remote sensing beamforming vector at the $k$-th satellite\\
$\mathbf{a}_t$ & Transmit steering vector of UPA \\
$\tau_{k,m}$, $f_{k,m}$ &  Time delay and Doppler deviation of the channel from the $k$-th LEO satellite to the $m$-th UE \\
${\alpha _{k,m}}$ & Navigation channel gain from the $k$-th LEO satellite to the $m$-th UE\\
$\theta_{k,m}$, $\varphi_{k,m}$ & Elevation and azimuth angle of the $m$-th UE relative to the UPA on the $k$-th LEO satellite\\
${\eta ^{RS}}$ & Minimum required SAINR threshold \\
$P_k^{\max}$ & Maximum transmit power budget of the $k$-th LEO satellite\\
$\mathbf{z}$ & Receive beamforming at the central satellite \\
${L}_k$ & Number of ambiguity areas of the $k$-th satellite\\
$\kappa$, $T$, $B$ & Boltzmann constant, noise temperature, bandwidth \\
$\lambda$, $f'$, $c$ & Wavelength, carrier frequency, speed of light \\
${\chi _{k,m}}$ & Rain attenuation factor \\
$b_{k,m}$ & Satellite array gain \\
\hline
\end{tabular}
\label{tab:notation}
\end{table}

\begin{figure}[!t] \centering
\includegraphics [width=0.49\textwidth] {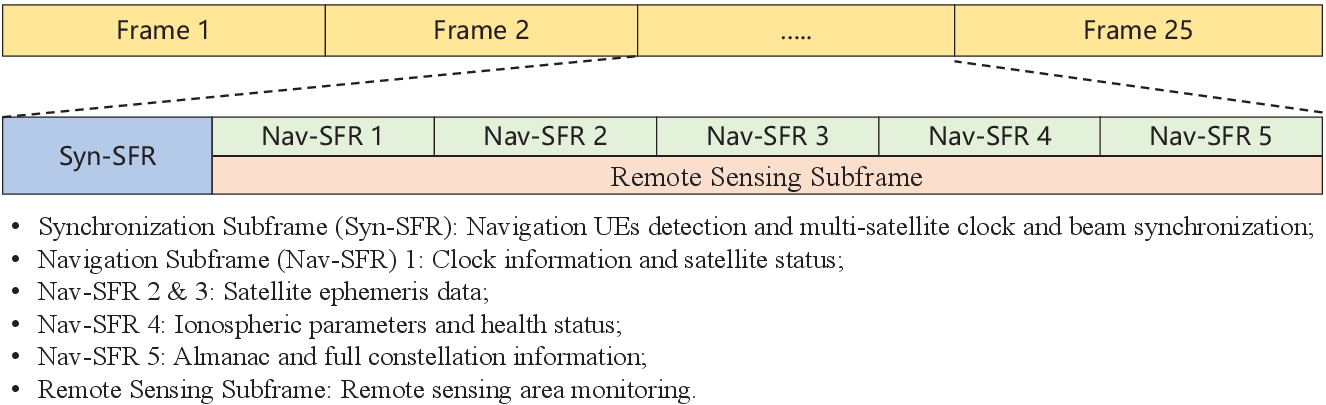}
\caption {Frame structure of integrated navigation and remote sensing signals in LEO satellite constellations.}
\label{frame}
\vspace{-10pt}
\end{figure}

To effectively achieve the dual functions, we design an unified signal frame structure for integrated navigation and remote sensing, referencing the famous GPS navigation message signals \cite{GPS Navigation Message}, as illustrated in Fig. \ref{frame}. At the beginning of each frame, the synchronization subframe (Syn-SFR) is used for UE detection as well as clock and beamforming synchronization across the whole service group via inter-satellite links. Then, data subframe is used to transfer both navigation and remote sensing information in the same spectrum\footnote{The integration of navigation and remote sensing over the same spectrum improves spectral efficiency, but also introduces mutual interference risks. To address this, the proposed system employs spatial-domain isolation through beamforming to separate navigation and sensing signals. Additionally, code-domain separation using orthogonal pseudo-random sequences is applied among navigation UEs to suppress inter-user interference. Future extensions may adopt time-domain separation if stricter isolation is required.}.
In this context, the $k$-th satellite transmits a dual-function signal for data subframe at the time index $t$ as
\begin{equation}\label{transmit signal}
{{\bf{x}}_k(t)} = {{\bf{w}}_k}s_k^{{\rm{RS}}}(t) + \sum\limits_{m = 1}^M {{{\bf{v}}_{k,m}}s_{k,m}^{{\rm{NAV}}}(t)},
\end{equation}
where ${{\bf{w}}_k} \in {\mathbb{C}^{N \times 1}}$ denotes the remote sensing beamforming vector at the $k$-th satellite for shaping the remote sensing signal $s_k^{{\rm{RS}}}$, while ${{\bf{v}}_{k,m}} \in {\mathbb{C}^{N \times 1}}$ denotes the navigation beamforming vector at the $k$-th satellite for delivering the pseudo-random coded navigation signal $s_{k,m}^{{\rm{NAV}}}$ to the $m$-th UE. Herein, $s_k^{\text{RS}}$ and $s_{k,m}^{\text{NAV}}$ are both assumed to be unit-power signals and are treated as statistically independent under an equivalent second-order statistical modeling framework, where these deterministic signals are approximated as zero-mean complex Gaussian processes to enable tractable interference and performance analysis. Hence, the average transmit power of the $k$-th satellite is given by
\begin{equation}\label{transmit signal power}
{P_k} = {\left\| {{{\bf{w}}_k}} \right\|^2} + \sum\limits_{m = 1}^M {{{\left\| {{{\bf{v}}_{k,m}}} \right\|}^2}}.
\end{equation}

Next, we will sequentially introduce both the navigation and remote sensing models in detail, along with their respective performance metrics.

\subsection{Navigation Model}
\begin{figure}[!t] \centering
\includegraphics [width=0.5\textwidth] {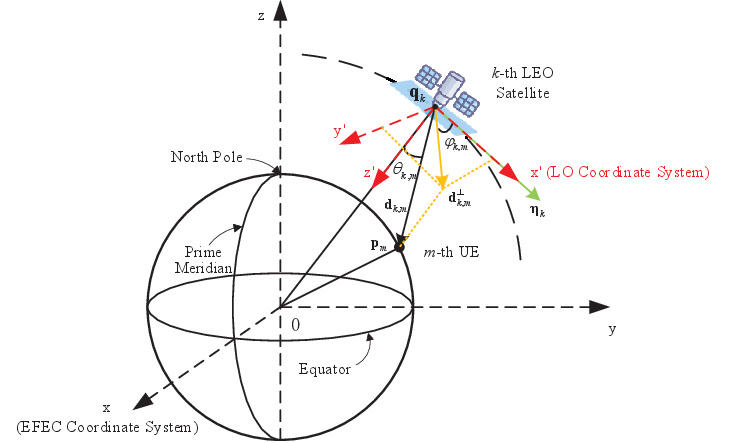}
\caption {Geometric relationships between LEO satellites and UEs in the ECEF and LO coordinate systems.}
\label{angle}
\end{figure}

For the navigation model, geometric relationships between LEO satellites and UEs play a crucial role, as shown in Fig. \ref{angle}. To facilitate the analysis and description of the navigation process, an Earth-Centered Earth-Fixed (ECEF) coordinate system is introduced\footnote{In the ECEF coordinate system, the origin is set at the Earth's center, the z-axis aligns with the Earth's rotational axis and points toward the North Pole, the x-axis points to the intersection of the prime meridian and the equatorial plane, and the y-axis is perpendicular to the xOz plane, forming a right-handed coordinate system.}. Within this ECEF coordinate system, the position of the $k$-th satellite is represented as ${{\bf{q}}_k} = \left( {q_k^x;q_k^y;q_k^z} \right)$, and its velocity is given by ${{\bm{\eta }}_k} = \left( {\eta _k^x;\eta _k^y;\eta _k^z} \right)$. These parameters are available to the UEs through regularly broadcast ephemeris data in standard satellite navigation systems. Likewise, the position and velocity of the $m$-th UE are denoted as ${{\bf{p}}_m} = \left( {p_m^x;p_m^y;p_m^z} \right)$ and ${{\bm{\gamma }}_m} = \left( {\gamma _m^x;\gamma _m^y;\gamma _m^z} \right)$, respectively. In addition, a Local Orbital (LO) coordinate system is defined for each LEO satellite. Specifically, the LO coordinate system for the $k$-th LEO satellite is centered at the satellite itself. The unit $\text{z}'_k$-axis is aligned with the normal of the UPA pointing toward the Earth's center, expressed as ${\bf{z}}'_k = \frac{{ - {{\bf{q}}_k}}}{{\left\| {{{\bf{q}}_k}} \right\|}} = \frac{{\left( { - q_k^x, - q_k^y, - q_k^z} \right)}}{{\sqrt {{{\left( {q_k^x} \right)}^2} + {{\left( {q_k^y} \right)}^2} + {{\left( {q_k^z} \right)}^2}} }}$. The unit $\text{x}'_k$-axis aligns with the direction of the satellite's orbital velocity, given by ${\mathbf{x}'_k} = \frac{{{{\bm{\eta }}_k}}}{{\left\| {{{\bm{\eta }}_k}} \right\|}} =  \frac{{\left( {\eta _k^x;\eta _k^y;\eta _k^z} \right)}}{{\sqrt {{{\left( {\eta _k^x} \right)}^2} + {{\left( {\eta _k^y} \right)}^2} + {{\left( {\eta _k^z} \right)}^2}} }}$. The unit $\text{y}'_k$-axis is orthogonal to both the $\text{x}'_k$-axis and $\text{z}'_k$-axis, forming a right-handed coordinate system, which is calculated as ${{\bf{y}}'_k} = \frac{{{\bf{z}}'_k \times {{\mathbf{x}}'_k}}}{{\left\| {{\bf{z}}'_k \times {{\mathbf{x}}'_k}} \right\|}}$.
Based on the above coordinate systems, the elevation angle ${\theta _{k,m}}$ and azimuth angle ${\varphi _{k,m}}$ of the $m$-th UE relative to the geocentric pointing UPA on the $k$-th LEO satellite can be respectively determined as
\begin{equation}\label{elevation angle}
{\theta _{k,m}} = \arccos \frac{{{{\bf{d}}_{k,m}} \cdot {{{\bf{z}}}'_k}}}{{\left\| {{{\bf{d}}_{k,m}}} \right\| \cdot \left\| {{{{\bf{z}}}'_k}} \right\|}} \in \left[ {0,\frac{\pi }{2}} \right],
\end{equation}
\begin{equation}\label{azimuth angle}
{\varphi _{k,m}} = {\rm{atan2}}\left( {{\bf{d}}_{k,m}^ \bot  \cdot {{{\bf{y}}}'_k},{\bf{d}}_{k,m}^ \bot  \cdot {{\mathbf{x}}'_k}} \right) \in \left[ { - \pi ,\pi } \right],
\end{equation}
where ${{\bf{d}}_{k,m}} = {{\bf{p}}_m} - {{\bf{q}}_k}$ denotes the direction vector from the $k$-th satellite to the $m$-th UE, and ${\bf{d}}_{k,m}^ \bot  = {\bf{d}}_{k,m}^{} - \big( {{\bf{d}}_{k,m}^{} \cdot {\bf{z}}'_k} \big){\bf{z}}'_k$ denotes the projection vector of the direction vector ${{\bf{d}}_{k,m}}$ onto the plane orthogonal to the $\text{z}'_k$-axis.

In this context, the received signal at the $m$-th UE through the satellite-terrestrial channel can be expressed as
\begin{align}\label{received signal NUE}
y_m^{{\rm{r}}}(t) = &\sum\limits_{k = 1}^K {{\alpha _{k,m}}{\bf{a}}_t^{\rm H}\left( {{\theta _{k,m}},{\varphi _{k,m}}} \right){{\bf{x}}_k}\left( {t - {\tau _{k,m}}} \right)e^{ {j2\pi {f_{{{k,m}}}}t} }}  \notag\\
&\ \ + {n_m}\left( t \right),
\end{align}
where ${\tau _{k,m}}$ and ${f_{k,m}}$ are the time delay and Doppler deviation of the channel from the $k$-th LEO satellite to the $m$-th UE, respectively\footnote{It is crucial to clarify that in this received signal model, ${\tau _{k,m}}$ represents the true geometric propagation time delay of the signal from the $k$-th satellite to the $m$-th UE, reflecting the physical time taken for the signal to travel this distance. This delay is defined with respect to the satellite clock, consistent with how GNSS signals are time-tagged at the satellite side. It is a component of the physical signal model. While real-world systems inherently involve clock biases at both the satellite and the UE, these are not assumed to be perfectly synchronized or pre-compensated at this signal modeling stage. Instead, their effects are explicitly accounted for and estimated as unknown parameters within the subsequent pseudo-range measurement and PVT estimation algorithm.}. Herein, ${f_{k,m}}$ is related to velocity ${{\bm{\gamma }}_m}$ and can be expressed as ${f_{k,m}} =  - {\left( {{{\bm{\eta }}_k} - {{\bm{\gamma }}_m}} \right)^{\rm T}}{{\bf{u}}_{k,m}}\frac{{f'}}{c}$ with ${{\bf{u}}_{k,m}} = \frac{{{{\bf{q}}_k} - {{\bf{p}}_m}}}{{\left\| {{{\bf{q}}_k} - {{\bf{p}}_m}} \right\|}}$, $f'$ denoting the signal frequency, and $c$ denoting the speed of light. Besides, ${n_m}\left( t \right)$ denotes the additive white Gaussian noise (AWGN) with variance $\sigma_m^2=\kappa BT$ received at the $m$-th UE, and $\alpha _{k,m}$ denotes the downlink channel gain from the $k$-th LEO satellite to the $m$-th UE given by \cite{satellite-terrestrial channel}
\begin{equation}\label{channel gain}
{\alpha _{k,m}} = \sqrt {{{\left( {\frac{c}{{4\pi f'\left\| {{{\bf{d}}_{k,m}}} \right\|}}} \right)}^2}{{{G_m}}}}  \cdot {\chi _{k,m}} \cdot \sqrt {b_{k,m}},
\end{equation}
where $\kappa$, $T$, $B$, and $G_m$ are Boltzmann constant, noise temperature, channel bandwidth, and receive antenna gain, respectively. The rain attenuation factor is expressed as ${\chi _{k,m}} = {\xi_r ^{\frac{1}{2}}}{e^{ - j{\psi _{k,m}}}}$, where the rain attenuation gain ${\xi_r ^{\frac{1}{2}}}$ follows a complex Gaussian distribution $\mathcal{C}\mathcal{N}\left( {{\mu _r},\sigma _r^2} \right)$, and ${\psi _{k,m}}$ represents the phase vector between the $k$-th LEO satellite and the $m$-th UE. Moreover, the satellite array gain $b_{k,m}$ is given by $b_{k,m} = b_k^{\max }{\left( {{{{J_1(u')}}}/{{(2u')}} + 36{{{J_3(u')}}}/{{{(u'^3)}}}} \right)^3}$, where $u' = 2.071\left( {\sin \left( {{\varepsilon _{k,m}^b}} \right)/\sin \left( {\varepsilon _k^{3{\rm{dB}}}} \right)} \right)$ with ${\varepsilon _{k,m}^{b}}$ being the antenna angle between the $k$-th LEO satellite and the $m$-th UE, and $b_k^{\max }$ and ${\varepsilon _k^{3{\rm{dB}}}}$ denoting the maximum satellite array gain and 3-dB angle of the $k$-th LEO satellite, respectively \cite{beam gain}. Herein, $J_1(\cdot)$ and $J_3(\cdot)$ denote the first-order and third-order Bessel functions, respectively. In addition, ${\bf{a}}_t\left( {{\theta _{k,m}},{\varphi _{k,m}}} \right)$ denotes the transmit steering vector, which is expressed as equation (\ref{steering vector}) at the top of next page
\begin{figure*}[!t]
\begin{equation}\label{steering vector}
{{\bf{a}}_t}\left( {{\theta _{k,m}},{\varphi _{k,m}}} \right) = \frac{1}{{\sqrt {N} }}
\left[ \begin{array}{l}
1; \cdots ;{e^{j\frac{{2\pi }}{\lambda }d\left( {{i_x}\cos {\varphi _{k,m}}\sin {\theta _{k,m}} + {i_y}\sin {\varphi _{k,m}}\sin {\theta _{k,m}}} \right)}}; \cdots \\
;{e^{j\frac{{2\pi }}{\lambda }d\left( {\left( {{N_x} - 1} \right)\cos {\varphi _{k,m}}\sin {\theta _{k,m}} + \left( {{N_y} - 1} \right)\sin {\varphi _{k,m}}\sin {\theta _{k,m}}} \right)}}
\end{array} \right],
\end{equation}
\hrulefill
\end{figure*}
Herein, $N = {N_x}{N_y}$ represents the size of the UPA, with $N_x$ and $N_y$ denoting the number of antennas in the $\text{x}'_k$ and $\text{y}'_k$ directions, respectively, $d$ is the antenna spacing, $\lambda$ is the signal wavelength, and $i_x$ and $i_y$ are the indexes of the antennas along the $\text{x}'_k$ and $\text{y}'_k$ directions, respectively.

Subsequently, the $m$-th UE decodes the received signal $y_m^{{\rm{r}}}$ by using its pseudo-random code for navigation, and the decoded signal associated with the $k$-th satellite can be expressed as
\begin{align}\label{pseudo-random code decoding}
&y_{k,m}^{{\rm{d}}}(t)
{=} {{\alpha _{k,m}}{\bf{a}}_t^{\rm H}\left( {{\theta _{k,m}},{\varphi _{k,m}}} \right){{\bf{v}}_{k,m}}\tilde{s}_{k,m}^{{\rm{NAV}}}\left( {t - {\tau _{k,m}}} \right)e^{ {j2\pi {f_{{{k,m}}}}{t}} }}  \notag\\
&\ \ \ \ \ \ \ \ + {i_m^d}\left( t \right)+ {n_m^d}\left( t \right),
\end{align}
where $\tilde{s}_{k,m}^{{\rm{NAV}}}\left( t- {\tau _{k,m}} \right)$ is the navigation data signal obtained by decoding ${s}_{k,m}^{{\rm{NAV}}}\left( t- {\tau _{k,m}} \right)$,
${i_m^d(t)}$ is the pseudo-random spreading signal of the remote sensing interference ${i_m(t)} = \sum\limits_{r = 1}^K {{\alpha _{r,m}}{\bf{a}}_t^{\rm H}\left( {{\theta _{r,m}},{\varphi _{r,m}}} \right){{\bf{w}}_r}s_r^{{\rm{RS}}}\left( t- {\tau _{r,m}} \right)}$, and ${n_m^d}\left( t \right)$ is the pseudo-random spreading signal of ${n_m}\left( t \right)$. Then, the decoded signal $y_{k,m}^{{\rm{d}}}$ is utilized for PVT. It is worth noting that in GNSS systems, there are various methods for obtaining PVT parameters, with the most common being the pseudo-range measurement method and the MLE method \cite{2SP_DPD}. The former method with pseudo-range measurements is relatively simple and involves lower computational complexity. However, it requires at least seven satellites and provides slightly lower accuracy. The latter one calculates PVT parameters directly from the received satellite signal data. This method fully leverages all available information to improve estimation accuracy but demands significant computing power at UEs due to its high time complexity.

\begin{figure}[!t] \centering
\centering
\includegraphics [width=0.4\textwidth] {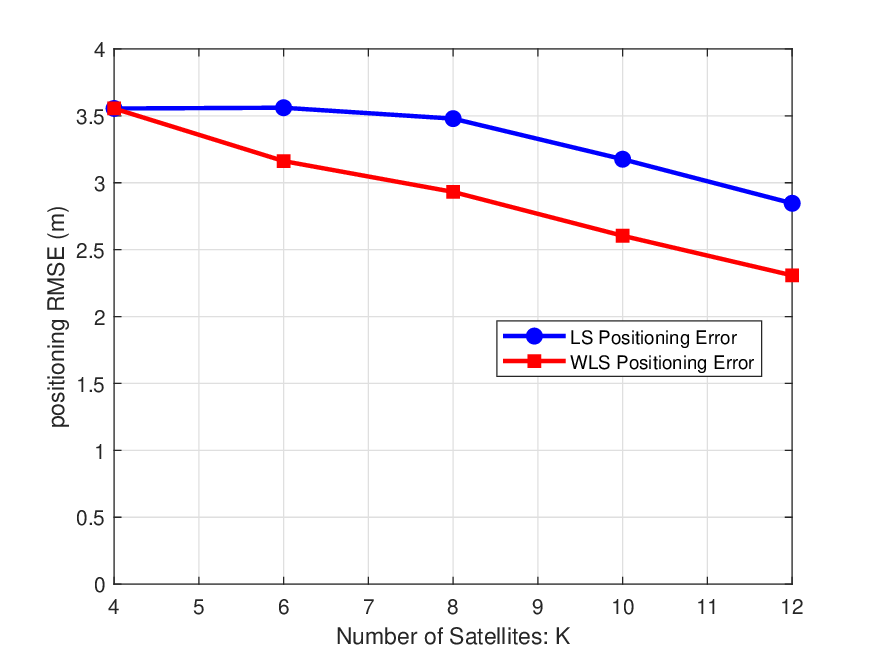}
\caption {Comparison of positioning performance for LS and WLS.}
\label{WLS}
\end{figure}

In this context, we propose a hybrid navigation method that combines the advantages of both the pseudo-range measurement and the MLE methods. Initially, the proposed method leverages pseudo-range information from multiple LEO satellites to ascertain the UE position and time error parameters. Subsequently, based on these preliminary results, it proceeds to estimate the velocity parameters through the MLE framework. By implementing above two steps, the proposed method can strike a balance between computational efficiency and the precision of PVT parameter estimation. Specifically, the pseudo-range measurement method calculates the distance between the UE and the satellite by measuring the propagation time between the signal transmission and reception, combined with the signal propagation speed, i.e., the pseudo-range ${\rho _{k,m}} = c{\hat \tau _{k,m}}$ with ${\hat \tau _{k,m}}$ being the time delay estimated based on equation (\ref{received signal NUE}). Therefore, pseudo-ranges establish a nonlinear relationship between the position of UE and the estimated time delay for each satellite as
\begin{equation}\label{pseudo-ranges relationship}
{\rho _{k,m}} = {\tilde d_{k,m}} + c\left( {\delta _m^t - \delta _k^{\rm{SAT}}} \right) + {\varepsilon _{k,m}},
\end{equation}
where ${\tilde d_{k,m}}=\|{{\bf{d}}_{k,m}}\|$ is the geometric distance from the $k$-th satellite to the $m$-th UE, $\delta _m^t$ is the unknown time error of the $m$-th UE, $\delta _k^{SAT}$ is the clock bias at the $k$-th LEO satellite, obtained in advance by UEs from the navigation information, and ${\varepsilon _{k,m}}$ is the pseudo-range error between the $k$-th satellite and the $m$-th UE. This error ${\varepsilon _{k,m}}$ primarily depends on the quality of the received navigation signals and is influenced by factors such as ionospheric and tropospheric delays, multipath biases, and noise-induced measurement errors \footnote{It is important to note that in LEO satellite links, particularly at the Ka-band frequencies employed here, the direct Line-of-Sight (LoS) path is strongly dominant. Therefore, multipath effects are primarily modeled as biases contributing to $\varepsilon_{k,m}$ rather than as significant fading, aligning with standard practices in high-precision satellite navigation \cite{footnote5_1}.}. Furthermore, uncertainties in the satellite position ${{\bf{q}}_k}$ due to ephemeris prediction errors or satellite clock drift also contribute to $\varepsilon_{k,m}$. In this case, the UE's position and time error are determined using the estimated pseudo-ranges from $K\left( {K \ge 4} \right)$ LEO satellites, which is given according to equation (\ref{pseudo-ranges relationship}) as
\begin{equation}\label{pseudo-ranges calculation}
{\rho _{k,m}} + c\delta _k^{SAT} - {\varepsilon _{k,m}} = {\tilde d_{k,m}} + c\delta _m^t, \ \ \forall k \in K.
\end{equation}
It is worth noting that equation (\ref{pseudo-ranges calculation}) is nonlinear, thus a first-order Taylor expansion is adopted for ${\tilde d_{k,m}}$ around the initial position estimation point ${\bf{p}}_m^{\left( 0 \right)} = \left( {p_m^{{x^{\left( 0 \right)}}}; p_m^{{y^{\left( 0 \right)}}}; p_m^{{z^{\left( 0 \right)}}}} \right)$ (obtained by using the Bancroft algorithm \cite{Bancroft}) as
\begin{equation}\label{pseudo-ranges Taylor}
{\tilde d_{k,m}} \approx \tilde{d}_{k,m}^{\left( 0 \right)} + \frac{{p_m^{{x^{\left( 0 \right)}}} - q_k^x}}{{\tilde{d}_{k,m}^{\left( 0 \right)}}}\delta _m^x + \frac{{p_m^{{y^{\left( 0 \right)}}} - q_k^y}}{{\tilde{d}_{k,m}^{\left( 0 \right)}}}\delta _m^y + \frac{{p_m^{{z^{\left( 0 \right)}}} - q_k^z}}{{\tilde{d}_{k,m}^{\left( 0 \right)}}}\delta _m^z,
\end{equation}
where $\tilde{d}_{k,m}^{\left( 0 \right)}=\left\| {{{\bf{q}}_k} - {\bf{p}}_m^{\left( 0 \right)}} \right\|$, $\delta _m^x = p_m^x - p_m^{{x^{\left( 0 \right)}}}$, $\delta _m^y = p_m^y - p_m^{{y^{\left( 0 \right)}}}$ and $\delta _m^z = p_m^z - p_m^{{z^{\left( 0 \right)}}}$.
Herein, the Bancroft algorithm provides a robust closed-form solution to initialization, ensuring a sufficiently accurate starting point for the subsequent weighted least squares (WLS) problem. This problem is typically solved through an iterative process, allowing for progressive refinement and convergence even if the initial estimate is not perfectly exact.
Specifically, substituting equation (\ref{pseudo-ranges Taylor}) into equation (\ref{pseudo-ranges calculation}) and introducing weighting matrix ${{\bf{\Phi }}_m}$, the PVT parameter estimation can be formulated as the following WLS problem:
\begin{equation}\label{eq_WLS}
{{\bm{\hat \delta }}_m} = \arg \mathop {\min }\limits_{{{\bm{\delta }}_m}} \left\{ {{{\left( {{{\bf{y}}_m} - {{\bf{Z}}_m}{{\bm{\delta }}_m}} \right)}^{\rm H}}{{\bf{\Phi }}_m}\left( {{{\bf{y}}_m} - {{\bf{Z}}_m}{{\bm{\delta }}_m}} \right)} \right\},
\end{equation}
where
\begin{equation}
{{\bf{y}}_m} = \left[ {\begin{array}{*{20}{c}}
{{\rho _{1,m}} + c\delta _1^{SAT} - {\varepsilon _{1,m}} - \tilde{d}_{1,m}^{\left( 0 \right)}}\\
 \vdots \\
{{\rho _{K,m}} + c\delta _K^{SAT} - {\varepsilon _{K,m}} - \tilde{d}_{K,m}^{\left( 0 \right)}}
\end{array}} \right],
\end{equation}
\begin{equation}
{{\bf{Z}}_m} = \left[ {\begin{array}{*{20}{c}}
{\frac{{p_m^{{x^{\left( 0 \right)}}} - q_1^x}}{{\tilde{d}_{1,m}^{\left( 0 \right)}}}}&{\frac{{p_m^{{y^{\left( 0 \right)}}} - q_1^y}}{{\tilde{d}_{1,m}^{\left( 0 \right)}}}}&{\frac{{p_m^{{z^{\left( 0 \right)}}} - q_1^z}}{{\tilde{d}_{1,m}^{\left( 0 \right)}}}}&c\\
 \vdots & \vdots & \vdots & \vdots \\
{\frac{{p_m^{{x^{\left( 0 \right)}}} - q_K^x}}{{\tilde{d}_{K,m}^{\left( 0 \right)}}}}&{\frac{{p_m^{{y^{\left( 0 \right)}}} - q_K^y}}{{\tilde{d}_{K,m}^{\left( 0 \right)}}}}&{\frac{{p_m^{{z^{\left( 0 \right)}}} - q_K^z}}{{\tilde{d}_{K,m}^{\left( 0 \right)}}}}&c
\end{array}} \right],
\end{equation}
\begin{equation}
{{\bm{\delta }}_m} = \left[ {\delta _m^x;\delta _m^y;\delta _m^z;\delta _m^t} \right].
\end{equation}
In particular, to reflect heterogeneous measurement reliability across satellites in a low-complexity and sensor-agnostic manner, we set the observation weight for each pseudo-range to increase monotonically with the elevation angle of the corresponding satellite relative to the UE. Based on this, the weighting matrix ${{\bf{\Phi }}_m}$ is constructed as
\begin{equation}
{{\bf{\Phi }}_m} = \frac{{{\rm{Diag}}\left( {{{\sin }}\left( {{{\tilde \theta }_{1,m}}} \right),{{\sin }}\left( {{{\tilde \theta }_{2,m}}} \right), \cdots ,{{\sin }}\left( {{{\tilde \theta }_{K,m}}} \right)} \right)}}{{\max \left\{ {{{\sin }}\left( {{{\tilde \theta }_{1,m}}} \right),{{\sin }}\left( {{{\tilde \theta }_{2,m}}} \right), \cdots ,{{\sin }}\left( {{{\tilde \theta }_{K,m}}} \right)} \right\}}},
\end{equation}
where the elevation angle ${\tilde \theta _{k,m}}$ of the $k$-th satellite with respect to the $m$-th UE is calculated as
\begin{equation}
{\tilde \theta _{k,m}} = \frac{\pi }{2} - \arccos \left( {\frac{{{\bf{p}}_m^{\left( 0 \right)} \cdot \left( {{{\bf{q}}_k} - {\bf{p}}_m^{\left( 0 \right)}} \right)}}{{\left\| {{\bf{p}}_m^{\left( 0 \right)}} \right\| \cdot \left\| {{{\bf{q}}_k} - {\bf{p}}_m^{\left( 0 \right)}} \right\|}}} \right) \in \left[ {0,\frac{\pi }{2}} \right].
\end{equation}
Clearly, the solution to the WLS problem (\ref{eq_WLS}) can be computed as
\begin{equation}\label{WLS reslut}
{{\bm{\hat \delta }}_m} = {\left( {{\bf{Z}}_m^{\rm H}{{\bf{\Phi }}_m}{\bf{Z}}_m^{}} \right)^{ - 1}}{\bf{Z}}_m^{\rm H}{{\bf{\Phi }}_m}{{\bf{y}}_m}.
\end{equation}
As a result, according to equation (\ref{WLS reslut}), the estimated position and time error of the $m$-th UE is given by
\begin{equation}\label{position and time error}
\left( {\begin{array}{*{20}{c}}
{{{{\bf{\hat p}}}_m}}\\
{{\hat \delta _m^t}}
\end{array}} \right) = \left( {\begin{array}{*{20}{c}}
{{\bf{p}}_m^{\left( 0 \right)}}\\
0
\end{array}} \right) + {\left( {{\bf{Z}}_m^{\rm H}{{\bf{\Phi }}_m}{{\bf{Z}}_m}} \right)^{ - 1}}{\bf{Z}}_m^{\rm H}{{\bf{\Phi }}_m}{{\bf{y}}_m}.
\end{equation}
Note that the elevation-based weighting matrix leverages the empirical fact that lower-elevation links experience longer atmospheric traversals and stronger multipath, leading to larger pseudo-range uncertainty \cite{WLS reference}. Increasing the weight with elevation therefore serves as a variance-aware approximation to the inverse error variances. We prefer elevation-based weights over channel-gain weighting because elevation is geometry-derived, stable across satellites and time, and provides a robust low-complexity proxy without per-link calibration. We adopt WLS in (\ref{eq_WLS}) to explicitly accommodate variance heterogeneity across satellites. Under an additional Gaussian assumption, (\ref{eq_WLS}) coincides with the MLE solution. Fig. \ref{WLS} further shows that, under heterogeneous measurement qualities, the WLS method consistently outperforms the unweighted least squares (LS) baseline in positioning accuracy across different numbers of visible satellites.

Then, based on the estimated position ${{\bf{\hat p}}_m}$ and time error ${\hat \delta _m^t}$ parameters in equation (\ref{position and time error}), we use MLE to directly estimate the velocity parameter ${{\bm{\gamma }}_m}$ of the $m$-th UE.
Considering the decoded signal in \eqref{pseudo-random code decoding}, define the reference signal as
\begin{align}
u_{k,m}^{\text{ML}}\!\left(t;{\bm \gamma}_m\right)
\triangleq &{\bf a}_t^{H}\!\left(\theta_{k,m},\varphi_{k,m}\right)\,{\bf v}_{k,m}\,
\tilde s_{k,m}^{\rm NAV}\!\left(t-\tau_{k,m}\right)\, \notag\\
&\cdot{e^{j2\pi \left( { - {{\left( {{{\bm{\eta }}_k} - {{\bm{\gamma }}_m}} \right)}^{\rm{T}}}{{\bf{u}}_{k,m}}\frac{{f'}}{c}} \right){t}}}.
\end{align}
Thus, the decoded signal $y^{\rm d}_{k,m}(t)$ is given by
\begin{align}
y^{\rm d}_{k,m}(t)\;=\;\alpha_{k,m}\,u_{k,m}^{\text{ML}}\!\left(t;{\bm \gamma}_m\right)\;+\;i_m^{\rm d}(t)\;+\;n_m^{\rm d}(t).
\end{align}
Herein, the decoded noise $n_m^d(t)$ results from applying pseudo-random despreading to $n_m(t)$. Since despreading is a linear operation, the resulting decoded noise $n_m^d(t)$ preserves the Gaussian nature of the original noise, i.e., $n_m^d(t) \sim \mathcal{CN}(0, \sigma_m^2)$. The despread interference from remote sensing, formed as a weighted superposition of many contributors under typical LEO satellite constellations visibility and consistent with central limit considerations, is modeled as a zero mean complex Gaussian process whose second order statistics match those of the composite interference over long observation windows. Thus, the total noise and interference in the decoded signal can be modeled as Gaussian.
In this context, the MLE function of the unknown velocity parameter ${{\bm{\gamma }}_m}$ with fixed position ${{\bf{p}}_m}$ and time error ${\delta _m^t}$ can be given by
\begin{align}
\label{eq:joint_mle_alpha_gamma_twc}
\hat{\bm \gamma}_m^{\rm ML}
=\arg\min_{{\bm \gamma}_m}\ \min_{\alpha_{k,m}}
\sum_{k=1}^{K}\Big\|\,y^{\rm d}_{k,m}(\cdot)-\alpha_{k,m}\,u_{k,m}^{\text{ML}}(\cdot;{\bm \gamma}_m)\Big\|^{2},
\end{align}
where $\lVert x\rVert^{2}\!\triangleq\!\langle x,x\rangle$ and $\langle x,y\rangle$ is the standard $L_{2}$ inner product over the support of the known signal.
For any fixed ${\bm \gamma}_m$, the closed-form estimator of $\alpha_{k,m}$ is given by
\begin{align}
\label{eq:alpha_hat_twc}
\hat\alpha_{k,m}({\bm \gamma}_m)
=\frac{\big\langle u_{k,m}^{\text{ML}}(\cdot;{\bm \gamma}_m),\,y^{\rm d}_{k,m}(\cdot)\big\rangle}
{\big\|u_{k,m}^{\text{ML}}(\cdot;{\bm \gamma}_m)\big\|^{2}}.
\end{align}
Substituting \eqref{eq:alpha_hat_twc} into \eqref{eq:joint_mle_alpha_gamma_twc} and concentrating out ${\alpha_{k,m}}$, we have
\begin{align}
\label{eq:concentrated_gamma_twc}
\hat{\bm \gamma}_m^{\rm ML}
=\arg\max_{{\bm \gamma}_m}\ \sum_{k=1}^{K}
\frac{\big|\big\langle u_{k,m}^{\text{ML}}(\cdot;{\bm \gamma}_m),\,y^{\rm d}_{k,m}(\cdot)\big\rangle\big|^{2}}
{\big\|u_{k,m}^{\text{ML}}(\cdot;{\bm \gamma}_m)\big\|^{2}},
\end{align}
whose solution can be efficiently obtained by the particle swarm optimization (PSO) algorithm or the other fast grid search methods \cite{PSO_DPD}.

Next, we evaluate the performance of the proposed hybrid navigation algorithm in terms of the Gaussian-Equivalent CRB (GE-CRB). The GE-CRB represents the theoretical lower bound on the variance of unbiased estimators under the Gaussian-equivalent random-noise model, serving as a crucial benchmark for evaluating the accuracy of PVT multi-parameter estimation in navigation systems. For simplicity and consistency, we continue to refer to this bound as the CRB throughout the paper, with the understanding that it refers specifically to the GE-CRB derived under this modeling assumption. In particular, considering the received signal $y_{m}^{{\rm{r}}}$ in equation (\ref{received signal NUE}), we can obtain the equivalent received signal of all $K$ satellites at the $m$-th UE, which effectively eliminate inter-UE navigation interference by pseudo-random decoding and can be expressed as
\begin{align}\label{received signal NUE2}
&y_m^{{\rm{e}}}
{=} \sum\limits_{k = 1}^K {{\alpha _{k,m}}{\bf{a}}_t^{\rm H}\left( {{\theta _{k,m}},{\varphi _{k,m}}} \right){{\bf{v}}_{k,m}}s_{k,m}^{{\rm{NAV}}}\left( {t - {\tau _{k,m}}} \right)e^{ {j2\pi {f_{{{k,m}}}}t} }}  \notag\\
&\ \ \ \ \ + {i_m}\left( t \right)+ {n_m}\left( t  \right).
\end{align}
To further simplify the expression and facilitate the derivation, we define some intermediate variables as
\begin{align}
{\bf{g}}\left( {{{\bm{\theta }}_m},{{\bm{\varphi }}_m}} \right) =&[ {\bf{a}}_t^{\rm H}\left( {{\theta _{1,m}},{\varphi _{1,m}}} \right),{\bf{a}}_t^{\rm H}\left( {{\theta _{2,m}},{\varphi _{2,m}}} \right), \cdots ,\notag\\
&\ \ \ \ \ \ \ \text{\ \ }{\bf{a}}_t^{\rm H}\left( {{\theta _{K,m}},{\varphi _{K,m}}} \right) ] \in {\mathbb{C}^{1 \times NK}};\notag
\end{align}
\begin{equation}
\vspace{-2pt}
{{\bf{A}}_m} = {\rm{Diag}}\left( {{\alpha _{1,m}},{\alpha _{2,m}}, \cdots ,{\alpha _{K,m}}} \right) \otimes {{\bf{I}}_N} \in {\mathbb{C}^{NK \times NK}};\notag
\end{equation}
\begin{align}
\vspace{-2pt}
&{{\bf{D}}_m}\left( {t,{{\bm{\tau }}_m},{{\bf{f}}_{{m}}}} \right) = {\rm{Diag}}\big( s_{1,m}^{{\rm{NAV}}}\left( {t - {\tau _{1,m}}} \right) e^{j2\pi {f_{{{1,m}}}}t}, \notag\\
&\text{\ \ \ \ }s_{2,m}^{{\rm{NAV}}}\left( {t - {\tau _{2,m}}} \right) e^ {j2\pi {f_{{{2,m}}}}t}, \cdots ,s_{K,m}^{{\rm{NAV}}}\left( {t - {\tau _{K,m}}} \right) e^ {j2\pi {f_{{{K,m}}}}t} \big)\notag\\
&\text{\ \ \ \ }\otimes {{\bf{I}}_N} \in {\mathbb{C}^{NK \times NK}};\notag
\end{align}
\begin{equation}
\vspace{-2pt}
{{\bf{S}}^{{\rm{RS}}}} = {\rm{Diag}}\left( {s_1^{{\rm{RS}}},s_2^{{\rm{RS}}}, \cdots ,s_K^{{\rm{RS}}}} \right) \otimes {{\bf{I}}_N} \in {\mathbb{C}^{NK \times NK}}; \notag
\end{equation}
\begin{equation}
\vspace{-2pt}
{{\bf{\tilde v}}_m} = \left[ {{{\bf{v}}_{1,m}};{{\bf{v}}_{2,m}}; \cdots ;{{\bf{v}}_{K,m}}} \right] \in {\mathbb{C}^{NK \times 1}}; \notag
\end{equation}
\begin{equation}
\vspace{-2pt}
{\bf{\tilde w}} = \left[ {{{\bf{w}}_1};{{\bf{w}}_2}; \cdots ;{{\bf{w}}_K}} \right] \in {\mathbb{C}^{NK \times 1}}; \notag
\end{equation}
\begin{equation}
\vspace{-2pt}
{{\bm{\theta }}_m} = \left[ {{\theta _{1,m}};{\theta _{2,m}}; \cdots {\theta _{K,m}}} \right] \in {\mathbb{C}^{K \times 1}}; \notag
\end{equation}
\begin{equation}
\vspace{-2pt}
{{\bm{\varphi }}_m} = \left[ {{\varphi _{1,m}};{\varphi _{2,m}}; \cdots {\varphi _{K,m}}} \right] \in {\mathbb{C}^{K \times 1}}; \notag
\end{equation}
\begin{equation}
\vspace{-2pt}
{{\bm{\tau }}_m} = \left[ {{\tau _{1,m}};{\tau _{2,m}}; \cdots {\tau _{K,m}}} \right] \in {\mathbb{C}^{K \times 1}}; \notag
\end{equation}
\begin{equation}
\vspace{-2pt}
{{\bf{f}}_m} = \left[ {{f_{1,m}};{f_{2,m}}; \cdots {f_{K,m}}} \right] \in {\mathbb{C}^{K \times 1}}, \notag
\end{equation}
where ${\bf{\tilde w}}$ and ${{\bf{\tilde v}}_m}$ denote the equivalent  remote sensing and navigation beamforming vectors for LEO satellite constellations, respectively. Similarly, other newly defined symbols are considered jointly for $K$ LEO satellites. Herein, the time index of interference and noise is omitted to simplify the expression. Through such formal simplification, equation (\ref{received signal NUE2}) can be equivalently expressed as
\begin{equation}\label{received signal NUE3}
y_m^{{\rm{e}}} = {\mu _m} + {\tilde n_m},
\end{equation}
where ${\mu _m} = {\bf{g}}\left( {{{\bm{\theta }}_m},{{\bm{\varphi }}_m}} \right){{\bf{A}}_m}{{\bf{D}}_m}\left( {t,{{\bm{\tau }}_m},{{\bf{f}}_{{d_m}}}} \right){{\bf{\tilde v}}_m}$ denotes the useful signal, ${\tilde n_m} = {i_m} + {n_m}$ represents equivalent noise with variance $\tilde \sigma _m^2  = { \sigma _m^2} + {\left| {{\bf{g}}\left( {{{\bm{\theta }}_m},{{\bm{\varphi }}_m}} \right){{\bf{A}}_m}{\bf{\tilde w}}} \right|^2}$, and ${i_m} = {\bf{g}}\left( {{{\bm{\theta }}_m},{{\bm{\varphi }}_m}} \right){{\bf{A}}_m}{{\bf{S}}^{{\rm{RS}}}}{\bf{\tilde w}}$ denotes remote sensing interference. On the one hand, for the performance of position and time error estimation derived from pseudo-range measurements, we focus on the trace of the CRB matrix $\mathbf{C}_m$ for the $m$-th UE, which is given by \cite{CRB ref1}
\begin{equation}
{\rm{tr}}\left( {\bf{C}_m} \right) = {\rm{tr}}\left( {{{\left( {{\bf{F}}_{{{\bf{p}}_m},\delta _m^t}^{\rm{E}}} \right)}^{ - 1}}} \right){\rm{ = tr}}\left( {{{\bf{J}}_m}{{\left( {{\bf{F}}_{{{\bm{\tau }}_m}}^{\rm{E}}} \right)}^{ - 1}}{\bf{J}}_m^{\rm T}} \right),
\end{equation}
where ${{\bf{F}}_{{{\bf{p}}_m},\delta _m^t}^{\rm{E}}}$ denotes the Fisher information matrix (FIM) of the position and time error parameters $\left[ {{{{\bf{\hat p}}}_m};\hat \delta _m^t} \right]$, ${{\bf{F}}_{{{\bf{\tau }}_m}}^{\rm{E}}}$ represents the FIM of time delay ${{\bm{\tau }}_m}$, and ${{\bf{J}}_m} = c{\left( {{\bf{Z}}_m^{\rm H}{{\bf{\Phi }}_m}{{\bf{Z}}_m}} \right)^{ - 1}}{\bf{Z}}_m^{\rm H}{{\bf{\Phi }}_m}$ is the Jacobian matrix of the linear mapping relation from $\left[ {{{{\bf{p}}}_m};\delta _m^t} \right]$ to ${{\bm{\tau }}_m}$ according to equation (\ref{position and time error}). In order to derive the specific expression for the CRB from ${{\bf{F}}_{{{\bm{\tau }}_m}}^{\rm{E}}}$ in greater detail, we analyze the FIM with respect to the vector ${{\bm{\xi }}_m} = \left[ {{{\bm{\tau }}_m};{{\bm{f}}_{{m}}}} \right]$, which encompasses all unknown parameters, and is expressed as \cite{CRB ref2}
\begin{equation}
{\bf{F}}_m^{\bm{\xi }} = \left[ {\begin{array}{*{20}{c}}
{{\bf{F}}_{{{\bm{\tau }}_m}{{\bm{\tau }}_m}}^{}}&{{\bf{F}}_{{{\bm{\tau }}_m}{{\bf{f}}_{{m}}}}^{}}\\
{{\bf{F}}_{{{\bm{\tau }}_m}{{\bf{f}}_{{m}}}}^{\rm T}}&{{\bf{F}}_{{{\bf{f}}_{{m}}}{{\bf{f}}_{{m}}}}}
\end{array}} \right],
\end{equation}
where
\begin{equation}\label{FIM matrix}
{\left[ {{\bf{F}}_m^{\bm{\xi }}} \right]_{i,j}}
= \frac{2}{\tilde \sigma _m^2}{\mathop{\rm Re}\nolimits} \left\{ {\frac{{\partial \mu _m^{\rm H}}}{{\partial {{\left[ {{{\bm{\xi }}_m}} \right]}_i}}}\frac{{\partial \mu _m^{}}}{{\partial {{\left[ {{{\bm{\xi }}_m}} \right]}_j}}}} \right\}.
\end{equation}
The FIM in (28) follows from standard estimation theory for a deterministic signal in additive complex Gaussian noise \cite{CRB add}. This model is applicable to our case because the total disturbance $\tilde \sigma _m^2$ can be treated as a zero-mean complex Gaussian process. Specifically, the thermal noise $n_m(t)$ is inherently Gaussian, and the aggregate interference $i_m(t)$ is modeled as a zero mean complex Gaussian process whose second order statistics match those of the composite interference over long observation windows, a choice supported by typical LEO satellite constellations visibility with many contributors and consistent with central limit considerations.
In this context, we have \cite{CRB ref3}
\begin{equation}
{\bf{F}}_{{{\bm{\tau }}_m}}^{\rm{E}} = {\bf{F}}_{{{\bm{\tau }}_m}{{\bm{\tau }}_m}}^{} - {\bf{F}}_{{{\bm{\tau }}_m}{{\bf{f}}_{{m}}}}^{}{\bf{F}}_{{{\bf{f}}_{{m}}}{{\bf{f}}_{{m}}}}^{ - 1}{\bf{F}}_{{{\bm{\tau }}_m}{{\bf{f}}_{{m}}}}^{\rm T}.
\end{equation}
Then, we focus on the detailed expressions for ${\bf{F}}_{{{\bm{\tau }}_m}{{\bm{\tau }}_m}}^{}$, ${\bf{F}}_{{{\bf{f}}_{{m}}}{{\bf{f}}_{{m}}}}$ and ${\bf{F}}_{{{\bm{\tau }}_m}{{\bf{f}}_{{m}}}}$.
Specifically, the partial derivatives of ${\mu _m}$ with respect to the elements of unknown parameters ${{{\bm{\tau }}_m}}$ and ${{{\bf{f}}_{{m}}}}$ are determined as
\begin{equation}\label{mu_tau}
\frac{{\partial \mu _m^{}}}{{\partial {{\left[ {{{\bm{\tau }}_m}} \right]}_i}}} = {{\bf{g}}_m}{{\bf{A}}_m}\frac{{\partial {{\bf{D}}_m}}}{{\partial {{\left[ {{{\bm{\tau }}_m}} \right]}_i}}}{{\bf{\tilde v}}_m},
\end{equation}
with
\begin{align}
\frac{{\partial {{\bf{D}}_m}}}{{\partial {{\left[ {{{\bm{\tau }}_m}} \right]}_i}}} = {\rm{Diag}}\Big( 0,& \cdots , - \dot s_{i,m}^{{\rm{NAV}}}\left( {t - {\tau _{i,m}}} \right) \notag\\
& \exp \left\{ {j2\pi {f_{{{i,m}}}}t} \right\}, \cdots ,0 \Big) \otimes {{\bf{I}}_N},
\end{align}
and
\begin{equation}\label{mu_f}
\frac{{\partial \mu _m^{}}}{{\partial {{\left[ {{{\bf{f}}_{{m}}}} \right]}_i}}} = {{\bf{g}}_m}{{\bf{A}}_m}\frac{{\partial {{\bf{D}}_m}}}{{\partial {{\left[ {{{\bf{f}}_{{m}}}} \right]}_i}}}{{\bf{\tilde v}}_m},
\end{equation}
with
\begin{align}
\frac{{\partial {{\bf{D}}_m}}}{{\partial {{\left[ {{{\bf{f}}_{{m}}}} \right]}_i}}} = {\rm{Diag}}\Big( 0,& \cdots ,j2\pi ts_{i,m}^{{\rm{NAV}}}\left( {t - \tau _{i,m}} \right)\notag\\
&\exp \left\{ {j2\pi {f_{{{i,m}}}}t} \right\}, \cdots ,0 \Big) \otimes {{\bf{I}}_N},
\end{align}
where $\dot s_{i,m}^{{\rm{NAV}}} (t)$ denotes the time derivative of the navigation information signal $s_{i,m}^{{\rm{NAV}}} (t)$, and ${\bf{g}}_m$ and ${{\bf{D}}_m}$ are simplified forms of ${\bf{g}}\left( {{{\bm{\theta }}_m},{{\bm{\varphi }}_m}} \right)$ and ${{\bf{D}}_m}\left( {{{\bm{\gamma }}_m}} \right)$, respectively.
Then, the elements of the matrices ${{\bf{F}}_{{{\bm{\tau }}_m}{{\bm{\tau }}_m}}^{}}$, ${{\bf{F}}_{{{\bf{f}}_{_m}}{{\bf{f}}_{_m}}}^{}}$, and ${{\bf{F}}_{{{\bm{\tau }}_m}{{\bf{f}}_{_m}}}^{}}$ are expressed as
\begin{equation}\label{F_tautau_ele}
{\left[ {{\bf{F}}_{{{\bm{\tau }}_m}{{\bm{\tau }}_m}}^{}} \right]_{i,j}} = \frac{2}{{{{ {\tilde \sigma _m^2} }}}}{\mathop{\rm Re}\nolimits} \left\{ {{\bf{\tilde v}}_m^{\rm H}\frac{{\partial {\bf{D}}_m^{\rm H}}}{{\partial {{\left[ {{{\bm{\tau }}_m}} \right]}_i}}}{\bf{A}}_m^{\rm H}{\bf{g}}_m^{\rm H}{{\bf{g}}_m}{{\bf{A}}_m}\frac{{\partial {{\bf{D}}_m}}}{{\partial {{\left[ {{{\bm{\tau }}_m}} \right]}_j}}}{{{\bf{\tilde v}}}_m}} \right\},
\end{equation}
\begin{equation}\label{F_ff_ele}
{\left[ {{\bf{F}}_{{{\bf{f}}_{{m}}}{{\bf{f}}_{{m}}}}^{}} \right]_{i,j}} = \frac{2}{{{{ {\tilde \sigma _m^2} }} }}{\mathop{\rm Re}\nolimits} \left\{ {{\bf{\tilde v}}_m^{\rm H}\frac{{\partial {\bf{D}}_m^{\rm H}}}{{\partial {{\left[ {{{\bf{f}}_{{m}}}} \right]}_i}}}{\bf{A}}_m^{\rm H}{\bf{g}}_m^{\rm H}{{\bf{g}}_m}{{\bf{A}}_m}\frac{{\partial {{\bf{D}}_m}}}{{\partial {{\left[ {{{\bf{f}}_{{m}}}} \right]}_j}}}{{{\bf{\tilde v}}}_m}} \right\},
\end{equation}
\begin{equation}\label{F_tauf_ele}
{\left[ {{\bf{F}}_{{{\bm{\tau }}_m}{{\bf{f}}_{{m}}}}^{}} \right]_{i,j}} = \frac{2}{{{{ {\tilde \sigma _m^2} }} }}{\mathop{\rm Re}\nolimits} \left\{ {{\bf{\tilde v}}_m^{\rm H}\frac{{\partial {\bf{D}}_m^{\rm H}}}{{\partial {{\left[ {{{\bm{\tau }}_m}} \right]}_i}}}{\bf{A}}_m^{\rm H}{\bf{g}}_m^{\rm H}{{\bf{g}}_m}{{\bf{A}}_m}\frac{{\partial {{\bf{D}}_m}}}{{\partial {{\left[ {{{\bf{f}}_{{m}}}} \right]}_j}}}{{{\bf{\tilde v}}}_m}} \right\}.
\end{equation}
On the other hand, for velocity parameters estimation using MLE with obtained $\left[ {{{{\bf{p}}}_m};\delta _m^t} \right]$, we focus on the CRB matrix ${{\bf{C'_m}}}$ for the $m$-th UE, which is given by
\begin{equation}
{\rm{tr}}\left( {{\bf{C}'_m}} \right) = {\rm{tr}}\left( \left({{\bf{F}}_m^{\bf{\gamma }}}\right)^{-1} \right),
\end{equation}
where ${{\bf{F}}_m^{\bf{\gamma }}}$ is the FIM associated with the velocity parameter vector ${{{\bm{\gamma }}_m}}$, and its element is given by
\begin{align}\label{F_gamma_ele}
&{\left[ {{\bf{F}}_m^{\bf{\gamma }}} \right]_{i,j}}= \frac{2}{{{{ {\tilde \sigma _m^2} }}}}{\mathop{\rm Re}\nolimits} \left\{ {\frac{{\partial \mu _m^{\rm H}}}{{\partial {{\left[ {{{\bm{\gamma }}_m}} \right]}_i}}}\frac{{\partial \mu _m^{}}}{{\partial {{\left[ {{{\bm{\gamma }}_m}} \right]}_j}}}} \right\}\notag\\
&\ \ = \frac{2}{{{{ {\tilde \sigma _m^2} }}}}{\mathop{\rm Re}\nolimits} \left\{ {{\bf{\tilde v}}_m^{\rm H}\frac{{\partial {\bf{D}}_m^{\rm H}}}{{\partial {{\left[ {{{\bm{\gamma }}_m}} \right]}_i}}}{\bf{A}}_m^{\rm H}{\bf{g}}_m^{\rm H}{{\bf{g}}_m}{{\bf{A}}_m}\frac{{\partial {{\bf{D}}_m}}}{{\partial {{\left[ {{{\bm{\gamma }}_m}} \right]}_j}}}{{{\bf{\tilde v}}}_m}} \right\},
\end{align}
where
\begin{equation}
\frac{{\partial {\bf{D}}_m^{}}}{{\partial {{\left[ {{{\bm{\gamma }}_m}} \right]}_i}}}{\rm{ = Diag}}\left( {\left( {\sum\limits_{k = 1}^K {\frac{{\partial {\tilde{\bf{d}}_m}\left( {{{\bm{\gamma }}_m}} \right)}}{{\partial {{\left[ {{{\bf{f}}_{{m}}}} \right]}_k}}}} } \right) \odot \frac{{\partial {{\bf{f}}_{{m}}}}}{{\partial {{\left[ {{{\bm{\gamma }}_m}} \right]}_i}}}} \right) \otimes {{\bf{I}}_N},
\end{equation}
with
\begin{align}
{\tilde{\bf{d}}_m}\left( {{{\bm{\gamma }}_m}} \right) =\big[& s_{1,m}^{{\rm{NAV}}}\left( {t - {\tau _{1,m}}} \right) e^{ {j2\pi {f_{1,m}}\left( {{{\bm{\gamma }}_m}} \right)t} }; \cdots \notag\\
& \ \ \ \ \;s_{K,m}^{{\rm{NAV}}}\left( {t - {\tau _{K,m}}} \right) e^{ {j2\pi {f_{k,m}}\left( {{{\bm{\gamma }}_m}} \right)t} } \big],
\end{align}
\begin{equation}
\frac{{\partial {\tilde{\bf{d}}_m}\left( {{{\bm{\gamma }}_m}} \right)}}{{\partial {{\left[ {{{\bf{f}}_{{m}}}} \right]}_i}}} = \left[ {0; \cdots ;j2\pi ts_{i,m}^{{\rm{NAV}}}\left( {t - \tau _{i,m}^{}} \right) e^{ {j2\pi {f_{{{i,m}}}}t} }; \cdots ;0} \right],
\end{equation}
\begin{align}
\frac{{\partial {{\bf{f}}_{{m}}}}}{{\partial {{\left[ {{{\bm{\gamma }}_m}} \right]}_i}}} &= {\left[ {\frac{{\partial {{\bf{f}}_{{m}}}}}{{\partial {{\bm{\gamma }}_m}}}} \right]_{:,i}} = {\left[ {\frac{{\partial {f_{{{1,m}}}}}}{{\partial {{\bm{\gamma }}_m}}};\frac{{\partial {f_{{{2,m}}}}}}{{\partial {{\bm{\gamma }}_m}}}; \cdots ;\frac{{\partial {f_{{{K,m}}}}}}{{\partial {{\bm{\gamma }}_m}}}} \right]_{:,i}}\notag\\
&= {\left[ {{\bf{u}}_{1,m}^{\rm T}\frac{{{f'}}}{c};{\bf{u}}_{2,m}^{\rm T}\frac{{{f'}}}{c}; \cdots ;{\bf{u}}_{K,m}^{\rm T}\frac{{{f'}}}{c}} \right]_{:,i}}.
\end{align}
Based on the CRBs of the PVT parameter estimation, we present the performance metrics for the three key specific functionalities of the proposed hybrid navigation algorithm, i.e., positioning error $E_m^P$, timing error $E_m^T$, and velocity measurement error $E_m^V$ as follows
\begin{equation}\label{E_m^P}
E_m^P= {\rm{tr}}\left( {{\bf{\Lambda }}{{\bf{J}}_m}{{\left( {{\bf{F}}_{{{\bm{\tau }}_m}{{\bm{\tau }}_m}}^{} - {\bf{F}}_{{{\bm{\tau }}_m}{{\bf{f}}_{{m}}}}^{}{\bf{F}}_{{{\bf{f}}_{{m}}}{{\bf{f}}_{{m}}}}^{ - 1}{\bf{F}}_{{{\bm{\tau }}_m}{{\bf{f}}_{{m}}}}^{\rm T}} \right)}^{ - 1}}{{{{\bf{J}}_m}}^{\rm T}}{{\bf{\Lambda }}^{\rm T}}} \right),
\end{equation}
\begin{equation}\label{E_m^T}
E_m^T= {\tilde{\bm{\gamma}}}{\bf{J}}_m {\left( {{\bf{F}}_{{{\bm{\tau }}_m}{{\bm{\tau }}_m}}^{} - {\bf{F}}_{{{\bm{\tau }}_m}{{\bf{f}}_{{m}}}}^{}{\bf{F}}_{{{\bf{f}}_{{m}}}{{\bf{f}}_{{m}}}}^{ - 1}{\bf{F}}_{{{\bm{\tau }}_m}{{\bf{f}}_{{m}}}}^{\rm T}} \right)^{ - 1}}{\bf{J}}_m^{\rm T}{{\tilde{\bm{\gamma}}}^{\rm T}},
\end{equation}
\begin{equation}\label{E_m^V}
E_m^V = {\rm{tr}}\left( {{{\left( {{\bf{F}}_m^{\bf{\gamma }}} \right)}^{ - 1}}} \right),
\end{equation}
where constant matrix ${\bf{\Lambda }} = [1,0,0,0;0,1,0,0;0,0,1,0]$ and constant vector ${\tilde{\bm{\gamma}}} = \left[ {0,0,0,1} \right]$.
Equations (\ref{E_m^P})-(\ref{E_m^V}) reveal that the equivalent remote sensing beamforming ${{\bf{\tilde w}}}$ and equivalent navigation beamforming ${{{{\bf{\tilde v}}}_m}}$ for LEO satellite constellations are the key factors jointly influencing PVT performance metrics. Therefore, optimizing navigation performance through the joint design of appropriate ${{\bf{\tilde w}}}$ and ${{{{\bf{\tilde v}}}_m}}$ is effective.

\subsection{Remote Sensing Model}
\begin{figure}[!t] \centering
\vspace{-5pt}
\includegraphics [width=0.35\textwidth] {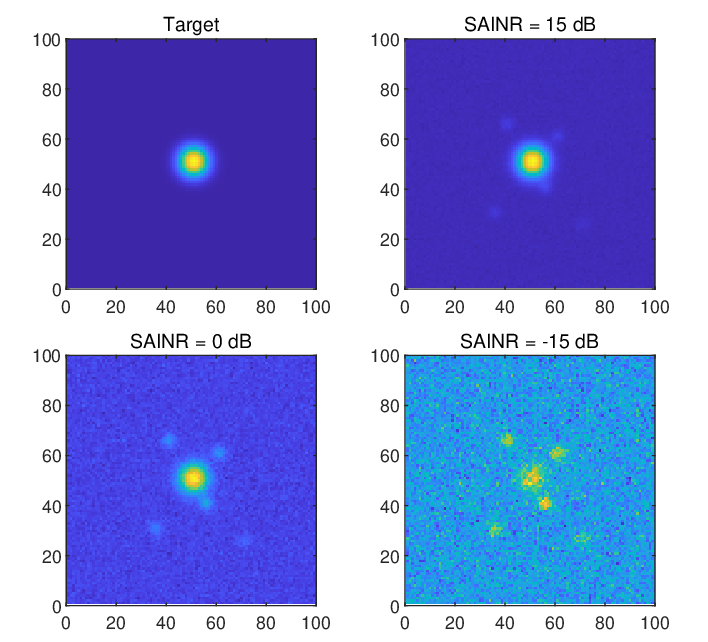}
\vspace{-7pt}
\caption {Remote sensing imaging results under different SAINR.}
\label{single_target}
\vspace{-15pt}
\end{figure}

For the remote sensing model, the received signal after time synchronization and Doppler compensation at the central satellite includes the desired signals reflected via the remote sensing area, ambiguity signals reflected via the ambiguity areas, navigation signal interference and noise, denoted as
\begin{align}
&{{\bf{y}}^{{\rm{RS}}}} = \underbrace {\sum\limits_{k = 1}^K {\beta {g'_k}{{\bf{a}}_r}\left( {{{\theta}'_1},{{\varphi}'_1}} \right){\bf{a}}_t^{\rm H}\left( {{{\theta}'_k},{{\varphi}'_k}} \right){{\bf{w}}_k}s_k^{{\rm{RS}}}} }_{{\text{desired signal}}}+\sum\limits_{k = 1}^K \sum\limits_{{l_{k = 1}}}^{{L_k}}
\notag\\
& \underbrace { {{{\tilde \beta }_{k,{l_k}}}{{\tilde g}_{k,{l_k}}}{{\bf{a}}_r}\left( {{{\tilde \theta '}_{1,k,{l_k}}},{{\tilde \varphi '}_{1,k,{l_k}}}} \right){\bf{a}}_t^{\rm H}\left( {{{\tilde \theta '}_{k,k,{l_k}}},{{\tilde \varphi '}_{k,k,{l_k}}}} \right){{\bf{w}}_k}s_k^{{\rm{RS}}}}  }_{{\text{ambiguity signal}}}\notag\\
&+ \underbrace {\sum\limits_{k = 1}^K {\sum\limits_{m = 1}^M {\beta {g'_k}{{\bf{a}}_r}\left( {{{\theta}'_1},{{\varphi}'_1}} \right){\bf{a}}_t^{\rm H}\left( {{{\theta}'_k},{{\varphi}'_k}} \right){{\bf{v}}_{k,m}}s_{k,m}^{{\rm{NAV}}}} } }_{{\text{navigation interference}}} + \underbrace {{{\bf{n}}^{{\rm{RS}}}}}_{{\text{noise}}},
\end{align}
where $\beta$ is the reflection coefficient of the remote sensing area, $g'_k$ is the round-trip channel gain between the $k$-th satellite and the remote sensing area, and $\theta'_1$ and $\varphi'_1$ are the elevation and azimuth angles from the central satellite to the remote sensing area, respectively. Similarly, $\theta'_k$ and $\varphi'_k$ are the elevation and azimuth angles from the $k$-th satellite to the remote sensing area, and ${{\bf{a}}_r}$ is the receive steering vector, as defined in equation (\ref{steering vector}). For the $k$-th satellite, $L_k$ represents the number of ambiguity areas, $l_k$ is the index of ambiguity areas, ${{\tilde \beta }_{k,{l_k}}}$ is the reflection coefficient of the $l_k$ remote sensing area, ${{\tilde g}_{k,{l_k}}}$ is the round-trip channel gain from the $k$-th satellite to the central satellite via reflection through the $l_k$ remote sensing area. Additionally, ${{\tilde \theta '}_{1,k,{l_k}}}$ and ${{\tilde \varphi '}_{1,k,{l_k}}}$ are the elevation and azimuth angles, respectively, from the central satellite to the $l_k$ ambiguity area of the $k$-th satellite, while ${{\tilde \theta '}_{k,k,{l_k}}}$ and ${{\tilde \varphi '}_{k,k,{l_k}}}$ represent the corresponding angles from the $k$-th satellite to the same ambiguity area. Finally, ${{{\bf{n}}^{{\rm{RS}}}}}\sim \mathcal{C}\mathcal{N}\left( {0,{{\sigma}}_s^2\mathbf{I}_N} \right)$ is AWGN received at the central satellite. Note that the angles mentioned above are determined based on the positions of the remote sensing area and ambiguity areas, which are calculated similar to equations (\ref{elevation angle}) and (\ref{azimuth angle}). Additionally, the symbols defined above refer to the central satellite when $k=1$. To suppress interference and enhance remote sensing performance, a receive beamforming with vector $\mathbf{z}$ is employed on the received signal ${{\bf{y}}^{{\rm{RS}}}}$ at the central satellite. Thus, the output of the receiver at the central satellite is represented as $r = {{\bf{z}}^{\rm H}}{{\bf{y}}^{{\rm{RS}}}}$. In this context, the signal-to-ambiguity-interference-noise ratio (SAINR) of the signal $r$ after applying receive beamforming is formulated as
\begin{equation}\label{SAINR}
\Gamma({\bf{z}}) = \frac{{{{\left| {{{\bf{z}}^{\rm H}}\beta {{\bf{a}}_r}\left( {{\theta' _1},{\varphi' _1}} \right){\bf{\tilde a}}_t^{}\left( {{\bm{\theta }'},{\bm{\varphi }'}} \right){\bf{G\tilde w}}} \right|}^2}}}{{{{\bf{z}}^{\rm H}}\left( {{\bf{R}} + {\sigma_s ^2}{{\bf{I}}_N}} \right){\bf{z}}}},
\end{equation}
where
\begin{equation}
{\bf{G}} = {\rm{Diag(}}{{{g}}'_1}{{,}}{{{g}}'_2}{{,}} \cdots {{,}}{{{g}}'_K}{{)}} \otimes {{\bf{I}}_N},\notag
\end{equation}
\begin{equation}
{\bm{\theta }'}=\left[{\theta' _1};{\theta' _2};\cdots;{\theta' _K} \right],\notag
\end{equation}
\begin{equation}
{\bm{\varphi }'}=\left[{\varphi' _1};{\varphi' _2};\cdots;{\varphi' _K} \right],\notag
\end{equation}
\begin{equation}
{\bf{\tilde a}}_t^{}\left( {{\bm{\theta }'},{\bm{\varphi }'}} \right) = [{\bf{a}}_t^{\rm H}\left( {{\theta' _1},{\varphi' _1}} \right),{\bf{a}}_t^{\rm H}\left( {{\theta' _2},{\varphi' _2}} \right), \cdots ,{\bf{a}}_t^{\rm H}\left( {{\theta' _K},{\varphi' _K}} \right)],\notag
\end{equation}
and the matrix notation ${\bf{R}}$ is defined by equation (\ref{R}) at the top of next page.
\begin{figure*}[!t]
\begin{align}\label{R}
{\bf{R}} =& \sum\limits_{k = 1}^K \sum\limits_{{l_{k = 1}}}^{{L_k}} {{\left| {{{\tilde \beta }_{k,{l_k}}}{{\tilde g}_{k,{l_k}}}} \right|}^2}{{\bf{a}}_r}\left( {{{\tilde \theta' }_{1,k,{l_k}}},{{\tilde \varphi' }_{1,k,{l_k}}}} \right){\bf{a}}_t^{\rm H}\left( {{{\tilde \theta' }_{k,k,{l_k}}},{{\tilde \varphi' }_{k,k,{l_k}}}} \right){{\bf{w}}_k}{\bf{w}}_k^{\rm H}{{\left( {{{\bf{a}}_r}\left( {{{\tilde \theta' }_{1,k,{l_k}}},{{\tilde \varphi' }_{1,k,{l_k}}}} \right){\bf{a}}_t^{\rm H}\left( {{{\tilde \theta' }_{k,k,{l_k}}},{{\tilde \varphi' }_{k,k,{l_k}}}} \right)} \right)}^{\rm H}}  \notag\\
&+ \sum\limits_{k = 1}^K {\sum\limits_{m = 1}^M {{{\left| {\beta {g'_k}} \right|}^2}{{\bf{a}}_r}\left( {{\theta' _1},{\varphi' _1}} \right){\bf{a}}_t^{\rm H}\left( {{\theta' _k},{\varphi' _k}} \right){{\bf{v}}_{k,m}}{\bf{v}}_{k,m}^{\rm H}{{\left( {{{\bf{a}}_r}\left( {{\theta' _1},{\varphi' _1}} \right){\bf{a}}_t^{\rm H}\left( {{\theta' _k},{\varphi' _k}} \right)} \right)}^{\rm H}}} }.
\end{align}
\end{figure*}
The optimal value of receive beamforming ${\bf{z}}$, which is derived by solving the SAINR maximization problem $\mathop {\max\  }\limits_{\bf{z}} \Gamma({\bf{z}})$ using the Minimum Variance Distortionless Response (MVDR) method, is explicitly provided in equation (\ref{z_solution}) \cite{MVDR}.
\begin{figure*}[!t]
\vspace{-13pt}
\begin{equation}\label{z_solution}
{{\bf{z}}^*} = \frac{{{{\left( {{\bf{R}} + {\sigma_s ^2}{{\bf{I}}_N}} \right)}^{ - 1}}{{\bf{a}}_r}\left( {{\theta' _1},{\varphi' _1}} \right){\bf{\tilde a}}_t^{}\left( {{\bm{\theta }'},{\bm{\varphi }'}} \right){\bf{G\tilde w}}}}{{{\bf{\tilde w}}{{\left( {{{\bf{a}}_r}\left( {{\theta' _1},{\varphi' _1}} \right){\bf{\tilde a}}_t^{}\left( {{\bm{\theta }'},{\bm{\varphi }'}} \right){\bf{G}}} \right)}^{\rm H}}{{\left( {{\bf{R}} + {\sigma_s ^2}{{\bf{I}}_N}} \right)}^{ - 1}}{{\bf{a}}_r}\left( {{\theta' _1},{\varphi' _1}} \right){\bf{\tilde a}}_t^{}\left( {{\bm{\theta }'},{\bm{\varphi }'}} \right){\bf{G\tilde w}}}}.
\end{equation}
\end{figure*}
Then, by substituting optimal solution ${{\bf{z}}^*}$ from equation (\ref{z_solution}) into equation (\ref{SAINR}), the maximum SAINR is derived and expressed in equation (\ref{max SAINR}).
\begin{figure*}[!t]
\vspace{-13pt}
\begin{equation}\label{max SAINR}
\Gamma\left( {{{\bf{z}}^*}} \right) = {\left| \beta  \right|^2}{{\bf{\tilde w}}^{\rm H}}{\left( {{{\bf{a}}_r}\left( {{\theta' _1},{\varphi' _1}} \right){\bf{\tilde a}}_t^{}\left( {{\bm{\theta }'},{\bm{\varphi }'}} \right){\bf{G}}} \right)^{\rm H}}{\left( {{\bf{R}} + {\sigma_s ^2}{{\bf{I}}_N}} \right)^{ - 1}}{{\bf{a}}_r}\left( {{\theta' _1},{\varphi' _1}} \right){\bf{\tilde a}}_t^{}\left( {{\bm{\theta }'},{\bm{\varphi }'}} \right){\bf{G\tilde w}}.
\end{equation}
\vspace{-13pt}
\hrulefill
\end{figure*}
The aforementioned equations (\ref{R})-(\ref{max SAINR}) are presented at the top of the next page.
In this context, the remote sensing performance is evaluated using the SAINR, which admits a closed-form expression and directly reflects the impact of ambiguity, interference, and noise. Unlike conventional SNR metrics that only account for thermal noise, the proposed SAINR explicitly captures the signal degradation caused by ambiguity signals (arising from reflections via unintended regions), co-channel interference from navigation signals (due to spectrum sharing), and background noise. Physically, the SAINR quantifies the reliability of the reflected sensing signal after receive beamforming at the central satellite, indicating how dominant the desired signal is relative to all undesired signal components. A higher SAINR means that the sensing target is well-isolated in the spatial and spectral domains, resulting in better detection, localization, and imaging quality. Moreover, key sensing metrics, such as imaging resolution and detection accuracy, are monotonic functions of SAINR. To intuitively demonstrate the impact of SAINR, we carry out an application using multiple-input multiple-output (MIMO)-SAR imaging technology \cite{MIMO_SAR}. As shown in Fig. \ref{single_target}, higher SAINR values for the received signal at the central satellite result in significantly improved accuracy in remote sensing imaging. From equation (\ref{max SAINR}), it is evident that the SAINR, serving as a performance metric for remote sensing, has an explicit closed-form relationship with the equivalent remote sensing beamforming ${{\bf{\tilde w}}}$ and equivalent navigation beamforming ${{{{\bf{\tilde v}}}_m}}$ for LEO satellite constellations. Consequently, it is feasible to improve the remote sensing performance also by designing appropriate ${{\bf{\tilde w}}}$ and ${{{{\bf{\tilde v}}}_m}}$.
It should be noted that the above analysis assumes perfect synchronization among cooperative satellites. However, in dynamic LEO environments, residual timing offsets, Doppler shifts, and clock mismatches are inevitable, resulting in asynchronous interference, signal misalignment, and SAINR degradation. To address this, the concept of an asynchronous factor has been introduced to model inter-satellite asynchrony and its impact on interference \cite{asynchronous}. This factor can be directly incorporated into the SAINR formulation by modifying interference terms to reflect time-domain misalignments, thereby enabling more accurate and robust beamforming design under practical conditions.

\section{Beamforming Design for Integrated Navigation and Remote Sensing}
Based on the previous derivation and analysis, it is known that both navigation beamforming and remote sensing beamforming are critical to the overall performance of the system. Thus, in this section, we present a beamforming design for integrated navigation and remote sensing in LEO satellite constellations by jointly optimizing navigation beamforming and remote sensing beamforming.

\subsection{Problem Formulation}
In order to improve the overall performance of the dual-function LEO satellite constellation, we propose to minimize the average of the weighted PVT error of all UEs, while ensuring compliance with the transmit power constraint as well as the SAINR requirements for the remote sensing signals at the central satellite. Mathematically, the design is formulated as the following optimization problem:
\begin{subequations}\label{OP1}
\begin{align}
\mathop {\min }\limits_{\tilde{\bf{ w}},{{\tilde{\bf{ v}}}_m}}\ \ & \frac{1}{M}\sum\limits_{m = 1}^M {\left( {{{{\rho_m ^P}}}E_m^P + {{{\rho_m ^T}}}E_m^T + {{{\rho_m ^V}}}E_m^V} \right)} \label{OP1obj} \\
{\rm{s.t.}}\ \ &{\left\| {{{\bf{\Delta }}_k}{\bf{\tilde w}}} \right\|^2} + \sum\limits_{m = 1}^M {{{\left\| {{{\bf{\Delta }}_k}{{{\bf{\tilde v}}}_m}} \right\|}^2}}  \le P_k^{\max }, \label{OP1st1} \\
&{\left| \beta  \right|^2}{{{\bf{\tilde w}}}^{\rm H}}{\left( {{{\bf{a}}_r}\left( {{\theta' _1},{\varphi' _1}} \right){\bf{\tilde a}}_t^{}\left( {{\bm{\theta }'},{\bm{\varphi }'}} \right){\bf{G}}} \right)^{\rm H}}{\left( {{\bf{R}} + {\sigma_s ^2}{{\bf{I}}_N}} \right)^{ - 1}}\cdot\notag\\
&\ \ \ \ {{\bf{a}}_r}\left( {{\theta' _1},{\varphi' _1}} \right){\bf{\tilde a}}_t^{}\left( {{\bm{\theta }'},{\bm{\varphi }'}} \right){\bf{G\tilde w}} \ge {\eta ^{RS}}, \label{OP1st2}
\end{align}
\end{subequations}
where ${{\bf{\Delta  }}_k} = {\bm{\varsigma }}_k \otimes {{\bf{I}}_N}\in {\mathbb{R}^{N \times NK}}$ with ${\bm{\varsigma }}_k$ being a $K$-dimensional row vector whose $k$-th element is set to 1 while all other elements are set to 0. Moreover, the objective function (\ref{OP1obj}) represents the weighted average of the PVT performance metrics outlined in equations (\ref{E_m^P})-(\ref{E_m^V}). Specifically, for the $m$-th UE, the weights are defined as $\rho_m ^P$ for positioning error, $\rho_m ^T$ for timing error, and $\rho_m ^V$ for velocity measurement error. Constraint (\ref{OP1st1}), derived from equation (\ref{transmit signal power}), imposes the transmit power limitations of LEO satellites, where $P_k^{\max}$ represents the maximum transmit power budget of the $k$-th LEO satellite. Meanwhile, constraint (\ref{OP1st2}) ensures the required quality of the remote sensing received signal at the central satellite under optimal receive beamforming, with ${\eta ^{RS}}$ denoting the minimum SAINR threshold required for remote sensing at the central satellite. This threshold is typically determined by the remote sensing application's performance requirements, such as the desired imaging resolution, detection probability, or target reconstruction clarity. It can be selected empirically through simulation studies or calibrated according to specific mission objectives \cite{SAINR threshold}.
Notably, the objective function and constraints of problem (\ref{OP1}) involve non-convex quadratic terms and complex matrix inversion operations with respect to the optimization variables ${\bf{\tilde w}}$ and ${{{\bf{\tilde v}}}_m}$. Thus, it is not feasible to obtain the optimal solution to problem (\ref{OP1}) within polynomial time. To tackle this challenge, we propose an efficient approach to derive a feasible suboptimal solution, aiming to enhance the performance of the integrated navigation and remote sensing system in LEO satellite constellations.

\subsection{Algorithm Design}
First of all, to address the non-convex objective function (\ref{OP1obj}), we introduce an auxiliary variable matrix ${{\bf{U}}_m} \in {\mathbb{C}^{K \times K}}$ for $E_m^P$ and $E_m^T$ with ${\bf{F}}_{{{\bm{\tau }}_m}{{\bm{\tau }}_m}}^{} - {\bf{F}}_{{{\bm{\tau }}_m}{{\bf{f}}_m}}^{}{\bf{F}}_{{{\bf{f}}_m}{{\bf{f}}_m}}^{ - 1}{\bf{F}}_{{{\bm{\tau }}_m}{{\bf{f}}_m}}^{\rm T}\succeq{{\bf{U}}_m}$. As a consequence, based on Schur complement theorem, the objective function can be reformulated as \cite{CRB ref2}
\begin{align}\label{OP2obj}
\mathop {\min }\limits_{{\bf{\tilde w}},{{{\bf{\tilde v}}}_m}, {\bf{U}}_m}& \frac{1}{M}\sum\limits_{m = 1}^M \Big( \rho _m^P{\rm{tr}}\left( {{\bf{\Lambda }}{{\bf{J}}_m}{\bf{U}}_m^{ - 1}{\bf{J}}_m^{\rm T}{{\bf{\Lambda }}^{\rm T}}} \right)+ \notag\\
&\ \ \rho _m^T{\tilde{\bm{\gamma}}}{{\bf{J}}_m}{\bf{U}}_m^{ - 1}{\bf{J}}_m^{\rm T}{{\tilde{\bm{\gamma}}}^{\rm T}}+ \rho _m^V{\rm{tr}}\left( {{{\left( {{\bf{F}}_m^{\bf{\gamma }}} \right)}^{ - 1}}} \right) \Big),
\end{align}
which is accompanied by the additional constraint
\begin{equation}\label{OP2st1}
\left[ {\begin{array}{*{20}{c}}
{{\bf{F}}_{{{\bm{\tau }}_m}{{\bm{\tau }}_m}}^{} - {{\bf{U}}_m}}&{{\bf{F}}_{{{\bm{\tau }}_m}{{\bf{f}}_m}}^{}}\\
{{\bf{F}}_{{{\bm{\tau }}_m}{{\bf{f}}_m}}^{\rm T}}&{{\bf{F}}_{{{\bf{f}}_m}{{\bf{f}}_m}}^{}}
\end{array}} \right]\succeq0.
\end{equation}
Next, to eliminate the matrix inversion operation in the objective function (\ref{OP2obj}), we introduce auxiliary variables ${{\bf{\Omega }}_m} \in {\mathbb{C}^{3 \times 3}}$ with ${{\bf{\Omega }}_m}\succeq{\bf{\Lambda }}{{\bf{J}}_m}{\bf{U}}_m^{ - 1}{\bf{J}}_m^{\rm T}{{\bf{\Lambda }}^{\rm T}}$, ${\Omega '_m} \in {\mathbb{C}^{1 \times 1}}$ with ${\Omega '_m} \ge {\tilde{\bm{\gamma}}}{\bf{J}}_m^{}{\bf{U}}_m^{ - 1}{\bf{J}}_m^{\rm T}{{\tilde{\bm{\gamma}}}^{\rm T}}$, and ${{\bf{\Omega ''}}_m} \in {\mathbb{C}^{3 \times 3}}$ with ${{\bf{\Omega ''}}_m}\succeq{\left( {{\bf{F}}_m^{\bf{\gamma }}} \right)^{ - 1}}$. Thus, the objective function (\ref{OP2obj}) is further simplified as
\begin{equation}\label{OP3obj}
\mathop {\min }\limits_{\scriptstyle{\bf{\tilde w}},{{{\bf{\tilde v}}}_m},{{\bf{U}}_m},\hfill\atop
\scriptstyle{{\bf{\Omega }}_m},{{\Omega}'_m},{{{\bf{\Omega }}}''_m}\hfill} \frac{1}{M}\sum\limits_{m = 1}^M {\left( {\rho _m^P{\rm{tr}}\left( {{{\bf{\Omega }}_m}} \right) + \rho _m^T{{\Omega '}_m} + \rho _m^V{\rm{tr}}\left( {{{{\bf{\Omega ''}}}_m}} \right)} \right)}.
\end{equation}
Similarly, by applying Schur complement theorem, the added constraints imposed on the auxiliary variables can be expressed as
\begin{equation}\label{OP3st1}
\left[ {\begin{array}{*{20}{c}}
{{{\bf{\Omega }}_m}}&{\mathbf{\Lambda} {\bf{J}}_m^{}}\\
{{\bf{J}}_m^{\rm T}{\mathbf{\Lambda} ^{\rm T}}}&{{{\bf{U}}_m}}
\end{array}} \right]\succeq0,
\end{equation}
\begin{equation}\label{OP3st2}
\left[ {\begin{array}{*{20}{c}}
{{{\Omega '}_m}}&{{\tilde{\bm{\gamma}}}{\bf{J}}_m^{}}\\
{{\bf{J}}_m^{\rm T}{{\tilde{\bm{\gamma}}}^{\rm T}}}&{{{\bf{U}}_m}}
\end{array}} \right]\succeq0,
\end{equation}
and
\begin{equation}\label{OP3st3}
\left[ {\begin{array}{*{20}{c}}
{{{{\bf{\Omega ''}}}_m}}&{{{\bf{I}}_3}}\\
{{{\bf{I}}_3}}&{{\bf{F}}_m^{\bf{\gamma }}}
\end{array}} \right]\succeq0.
\end{equation}
Unfortunately, matrix inequalities (\ref{OP2st1}) and (\ref{OP3st3}) remain non-convex due to the presence of quadratic terms in the elements of ${\bf{F}}_{{{\bm{\tau }}_m}{{\bm{\tau }}_m}}$, ${{\bf{F}}_{{{\bf{f}}_m}{{\bf{f}}_m}}^{}}$, ${{\bf{F}}_{{{\bm{\tau }}_m}{{\bf{f}}_m}}^{}}$, and ${{\bf{F}}_m^{\bf{\gamma }}}$. To solve this problem, we employ the semi-definite relaxation (SDR) technique and define ${\bf{W}} = {\bf{\tilde w}}{{\bf{\tilde w}}^{\rm H}}$ and ${{\bf{V}}_m} = {{\bf{\tilde v}}_m}{\bf{\tilde v}}_m^{\rm H}$. In this context, the elements of ${\bf{F}}_{{{\bm{\tau }}_m}{{\bm{\tau }}_m}}$, ${{\bf{F}}_{{{\bf{f}}_m}{{\bf{f}}_m}}^{}}$, ${{\bf{F}}_{{{\bm{\tau }}_m}{{\bf{f}}_m}}^{}}$, and ${{\bf{F}}_m^{\bf{\gamma }}}$ in equations
(\ref{F_tautau_ele})-(\ref{F_tauf_ele}) and (\ref{F_gamma_ele}) can be rewritten as
\begin{equation}\label{F_tautau_SDR}
{\left[ {{\bf{F}}_{{{\bm{\tau }}_m}{{\bm{\tau }}_m}}^{}} \right]_{i,j}} = \frac{2}{{{{ {\tilde \sigma _m^2} }}}}{\mathop{\rm Re}\nolimits} \left\{ {{{{\bf{V}}_m}}\frac{{\partial {\bf{D}}_m^{\rm H}}}{{\partial {{\left[ {{{\bm{\tau }}_m}} \right]}_i}}}{\bf{A}}_m^{\rm H}{\bf{g}}_m^{\rm H}{{\bf{g}}_m}{{\bf{A}}_m}\frac{{\partial {{\bf{D}}_m}}}{{\partial {{\left[ {{{\bm{\tau }}_m}} \right]}_j}}}} \right\},
\end{equation}
\begin{equation}\label{F_ff_SDR}
{\left[ {{\bf{F}}_{{{\bf{f}}_{{m}}}{{\bf{f}}_{{m}}}}^{}} \right]_{i,j}} = \frac{2}{{{{ {\tilde \sigma _m^2} }} }}{\mathop{\rm Re}\nolimits} \left\{ {{{{\bf{V}}_m}}\frac{{\partial {\bf{D}}_m^{\rm H}}}{{\partial {{\left[ {{{\bf{f}}_{{m}}}} \right]}_i}}}{\bf{A}}_m^{\rm H}{\bf{g}}_m^{\rm H}{{\bf{g}}_m}{{\bf{A}}_m}\frac{{\partial {{\bf{D}}_m}}}{{\partial {{\left[ {{{\bf{f}}_{{m}}}} \right]}_j}}}} \right\},
\end{equation}
\begin{equation}\label{F_tauf_SDR}
{\left[ {{\bf{F}}_{{{\bm{\tau }}_m}{{\bf{f}}_{{m}}}}^{}} \right]_{i,j}} = \frac{2}{{{{ {\tilde \sigma _m^2} }} }}{\mathop{\rm Re}\nolimits} \left\{ {{{{\bf{V}}_m}}\frac{{\partial {\bf{D}}_m^{\rm H}}}{{\partial {{\left[ {{{\bm{\tau }}_m}} \right]}_i}}}{\bf{A}}_m^{\rm H}{\bf{g}}_m^{\rm H}{{\bf{g}}_m}{{\bf{A}}_m}\frac{{\partial {{\bf{D}}_m}}}{{\partial {{\left[ {{{\bf{f}}_{{m}}}} \right]}_j}}}} \right\},
\end{equation}
\begin{equation}\label{F_gamma_SDR}
{\left[ {{\bf{F}}_m^{\bf{\gamma }}} \right]_{i,j}} = \frac{2}{{{{ {\tilde \sigma _m^2} }}}}{\mathop{\rm Re}\nolimits} \left\{ {{{\bf{V}}_m}\frac{{\partial {\bf{D}}_m^{\rm H}}}{{\partial {{\left[ {{{\bm{\gamma }}_m}} \right]}_i}}}{\bf{A}}_m^{\rm H}{\bf{g}}_m^{\rm H}{{\bf{g}}_m}{{\bf{A}}_m}\frac{{\partial {{\bf{D}}_m}}}{{\partial {{\left[ {{{\bm{\gamma }}_m}} \right]}_j}}}} \right\},
\end{equation}
where ${\tilde \sigma _m^2}$ is similarly reexpressed as
${\tilde \sigma _m^2}={\sigma _m^2 + {\rm{tr}}\left\{ {{\bf{WA}}_m^{\rm H}{{\bf{g}}^{\rm H}}{\bf{g}}{{\bf{A}}_m}} \right\}}$. It is seen that variables ${\bf{W}}$ and ${{\bf{V}}_m}$ are coupled to each other in equations (\ref{F_tautau_SDR})-(\ref{F_gamma_SDR}). Therefore, we apply the successive convex approximation (SCA) method to these four equations at the first-order Taylor expansion point $\left( {{{\bf{W}}^{\#}},{\bf{V}}_m^{\#}} \right)$. The specific Taylor expansion formulation for ${\left[ {{\bf{F}}_{{{\bm{\tau }}_m}{{\bm{\tau }}_m}}^{}} \right]_{i,j}}$ is presented as (\ref{SCA_tautau}) at the top of next page. Similarly, ${\left[ {{\bf{F}}_{{{\bf{f}}_{{m}}}{{\bf{f}}_{{m}}}}^{}} \right]_{i,j}}$ ${\left[ {{\bf{F}}_{{{\bm{\tau }}_m}{{\bf{f}}_{{m}}}}^{}} \right]_{i,j}}$ and ${\left[ {{\bf{F}}_m^{\bf{\gamma }}} \right]_{i,j}}$ after using SCA method can be obtained by replacing ${\bf{\tilde A}}_m^1$ in equation (\ref{SCA_tautau}) with ${\bf{\tilde A}}_m^2$, ${\bf{\tilde A}}_m^3$, and ${\bf{\tilde A}}_m^4$, respectively.
\begin{figure*}[!t]
\begin{align}\label{SCA_tautau}
{\left[ {{\bf{F}}_{{{\bm{\tau }}_m}{{\bm{\tau }}_m}}^{}} \right]_{i,j}}\left( {{{\bf{V}}_m},{\bf{W}}} \right)
&= {\left[ {{\bf{F}}_{{{\bm{\tau }}_m}{{\bm{\tau }}_m}}^{}} \right]_{i,j}}\left( {{\bf{V}}_m^{\#},{{\bf{W}}^{\#}}} \right)+ {\rm{tr}}\left( {{{\left( {{\nabla _{{{\bf{V}}_m}}}{{\left[ {{\bf{F}}_{{{\bf{\tau }}_m}{{\bf{\tau }}_m}}^{}} \right]}_{i,j}}\left( {{\bf{V}}_m^{\#},{{\bf{W}}^{\#}}} \right)} \right)}^{\rm H}}\left( {{{\bf{V}}_m} - {\bf{V}}_m^{\#}} \right)} \right) \notag\\
&\ \ \ \ \ \ \ \ \ \ \ \ \ \ \ \ \ \ \ \ \ \ \ \ \ \ \ \ \ \ \ \ \ \ \ + {\rm{tr}}\left( {{{\left( {{\nabla _{\bf{W}}}{{\left[ {{\bf{F}}_{{{\bm{\tau }}_m}{{\bm{\tau }}_m}}^{}} \right]}_{i,j}}\left( {{\bf{V}}_m^{\#},{{\bf{W}}^{\#}}} \right)} \right)}^{\rm H}}\left( {{\bf{W}} - {{\bf{W}}^{\#}}} \right)} \right)\notag\\
&= \frac{{2{\rm{tr}}\left\{ {{\bf{V}}_m^{\#}{\bf{\tilde A}}_m^1} \right\}}}{{\sigma _m^2 + {\rm{tr}}\left\{ {{{\bf{W}}^{\#}}{\bf{\tilde B}}_m^{}} \right\}}} + \frac{{2{\rm{tr}}\left( {{\bf{\tilde A}}_m^1\left( {{{\bf{V}}_m} - {\bf{V}}_m^{\#}} \right)} \right)}}{{\sigma _m^2 + {\rm{tr}}\left\{ {{{\bf{W}}^{\#}}{\bf{\tilde B}}_m^{}} \right\}}} - \frac{{2{\rm{tr}}\left\{ {{\bf{V}}_m^{\#}{\bf{\tilde A}}_m^1} \right\}{\rm{tr}}\left( {{\bf{\tilde B}}_m^{}\left( {{\bf{W}} - {{\bf{W}}^{\#}}} \right)} \right)}}{{{{\left( {\sigma _m^2 + {\rm{tr}}\left\{ {{{\bf{W}}^{\#}}{\bf{\tilde B}}_m^{}} \right\}} \right)}^2}}},
\end{align}
\end{figure*}
Herein, we define ${\bf{\tilde B}}_m^{} = {\bf{A}}_m^{\rm H}{{\bf{g}}^{\rm H}}{\bf{g}}{{\bf{A}}_m}$, ${\bf{\tilde A}}_m^1 = \frac{{\partial {\bf{D}}_m^{\rm H}}}{{\partial {{\left[ {{{\bm{\tau }}_m}} \right]}_i}}}{\bf{A}}_m^{\rm H}{\bf{g}}_m^{\rm H}{{\bf{g}}_m}{{\bf{A}}_m}\frac{{\partial {{\bf{D}}_m}}}{{\partial {{\left[ {{{\bm{\tau }}_m}} \right]}_j}}}$, ${\bf{\tilde A}}_m^2 = \frac{{\partial {\bf{D}}_m^{\rm H}}}{{\partial {{\left[ {{{\bf{f}}_m}} \right]}_i}}}{\bf{A}}_m^{\rm H}{\bf{g}}_m^{\rm H}{{\bf{g}}_m}{{\bf{A}}_m}\frac{{\partial {{\bf{D}}_m}}}{{\partial {{\left[ {{{\bf{f}}_m}} \right]}_j}}}$,
${\bf{\tilde A}}_m^3 = \frac{{\partial {\bf{D}}_m^{\rm H}}}{{\partial {{\left[{{{\bm{\tau }}_m}}\right]}_i}}} {\bf{A}}_m^{\rm H}{\bf{g}}_m^{\rm H} {{\bf{g}}_m} \\
{{\bf{A}}_m}\frac{{\partial {{\bf{D}}_m}}}{{\partial {{\left[ {{{\bf{f}}_m}} \right]}_j}}}$, and
${\bf{\tilde A}}_m^4 = \frac{{\partial {\bf{D}}_m^{\rm H}}}{{\partial {{\left[ {{{\bm{\gamma }}_m}} \right]}_i}}}{\bf{A}}_m^{\rm H}{\bf{g}}_m^{\rm H}{{\bf{g}}_m}{{\bf{A}}_m}\frac{{\partial {{\bf{D}}_m}}}{{\partial {{\left[ {{{\bm{\gamma }}_m}} \right]}_j}}}$.
In this way, constraints (\ref{OP2st1}) and (\ref{OP3st3}) are eventually transformed into standard linear matrix inequalities (LMIs). Meanwhile, due to the introduction of ${\bf{W}}$ and ${{\bf{V}}_m}$, the transmit power constraint (\ref{OP1st1}) is rephrased as
\begin{equation}\label{power_SDR}
{\rm{tr}}\left( {{{\bf{\Delta }}_k}{\bf{W\Delta }}_k^{\rm T}{\rm{ + }}\sum\limits_{m = 1}^M {{{\bf{\Delta }}_k}{{\bf{V}}_m}{\bf{\Delta }}_k^{\rm T}} } \right) \le P_k^{\max }.
\end{equation}
Similarly, the SAINR requirement constraint (\ref{OP1st2}) for the central satellite using SDR is reformulated as
\begin{align}\label{SAINR_SDR}
&{\left| \beta  \right|^2} {\rm{tr}}\Big({\bf{W}} {\left( {{{\bf{a}}_r}\left( {{\theta' _1},{\varphi' _1}} \right){\bf{\tilde a}}_t^{}\left( {{\bm{\theta }'},{\bm{\varphi }'}} \right){\bf{G}}} \right)^{\rm H}}{\left( {{\bf{R}} + {\sigma_s ^2}{{\bf{I}}_N}} \right)^{ - 1}}\notag\\
&\ \ \ \ {{\bf{a}}_r}\left( {{\theta' _1},{\varphi' _1}} \right){\bf{\tilde a}}_t^{}\left( {{\bm{\theta }'},{\bm{\varphi }'}} \right){\bf{G}}\Big) \ge {\eta ^{RS}},
\end{align}
where notation $\mathbf{R}$ is also rewritten as equation (\ref{R_SDR}) at the top of next page.
\begin{figure*}[!t]
\vspace{-15pt}
\begin{align}\label{R_SDR}
{\bf{R}} =& \sum\limits_{k = 1}^K \sum\limits_{{l_{k = 1}}}^{{L_k}} {{\left| {{{\tilde \beta }_{k,{l_k}}}{{\tilde g}_{k,{l_k}}}} \right|}^2}{{\bf{a}}_r}\left( {{{\tilde \theta' }_{1,k,{l_k}}},{{\tilde \varphi' }_{1,k,{l_k}}}} \right){\bf{a}}_t^{\rm H}\left( {{{\tilde \theta' }_{k,k,{l_k}}},{{\tilde \varphi' }_{k,k,{l_k}}}} \right){{\bf{\Delta }}_k}{\bf{W\Delta }}_k^{\rm T}{{\left( {{{\bf{a}}_r}\left( {{{\tilde \theta' }_{1,k,{l_k}}},{{\tilde \varphi' }_{1,k,{l_k}}}} \right){\bf{a}}_t^{\rm H}\left( {{{\tilde \theta' }_{k,k,{l_k}}},{{\tilde \varphi' }_{k,k,{l_k}}}} \right)} \right)}^{\rm H}}  \notag\\
&+ \sum\limits_{k = 1}^K {\sum\limits_{m = 1}^M {{{\left| {\beta {g'_k}} \right|}^2}{{\bf{a}}_r}\left( {{\theta' _1},{\varphi' _1}} \right){\bf{a}}_t^{\rm H}\left( {{\theta' _k},{\varphi' _k}} \right){{{\bf{\Delta }}_k}{{\bf{V}}_m}{\bf{\Delta }}_k^{\rm T}}{{\left( {{{\bf{a}}_r}\left( {{\theta' _1},{\varphi' _1}} \right){\bf{a}}_t^{\rm H}\left( {{\theta' _k},{\varphi' _k}} \right)} \right)}^{\rm H}}} }.
\end{align}
\hrulefill
\end{figure*}
Hence, the original optimization problem (\ref{OP1}) is reexpressed as
\begin{subequations}\label{OP4}
\begin{align}
\mathop {\min }\limits_{\scriptstyle{\bf{W}},{{{\bf{V}}}_m},{{\bf{U}}_m},\hfill\atop
\scriptstyle{{\bf{\Omega }}_m},{{\Omega}'_m},{{{\bf{\Omega }}}''_m}\hfill} & \frac{1}{M}\sum\limits_{m = 1}^M {\left( {\rho _m^P{\rm{tr}}\left( {{{\bf{\Omega }}_m}} \right) + \rho _m^T{{\Omega '}_m} + \rho _m^V{\rm{tr}}\left( {{{{\bf{\Omega ''}}}_m}} \right)} \right)}\label{OP4obj}\\
{\rm{s.t.}}\ \ & (\ref{OP2st1}), (\ref{OP3st1}), (\ref{OP3st2}), (\ref{OP3st3}), (\ref{power_SDR}), (\ref{SAINR_SDR}), \notag \\
& {{\bf{W}}}\succeq0,  {\bf{V}}_m\succeq0, \label{OP4st1} \\
& {\rm{Rank}}\left( {{{\bf{W}}}} \right) = 1,  {\rm{Rank}}\left( {\bf{V}}_m \right) = 1, \label{OP4st2}
\end{align}
\end{subequations}
where variables $\mathbf{W}$ and $\mathbf{R}$ in non-convex constraint (\ref{SAINR_SDR}) exhibit a complex coupling relationship. To address this issue, we employ the block coordinate descent (BCD) method. Specifically, we first solve the optimization problem (\ref{OP4}) with fixed $\mathbf{R}$, then compute the updated value of $\mathbf{R}$ using equation (\ref{R_SDR}), and incorporate updated $\mathbf{R}$ into the next iteration. Moreover, the application of SDR technique introduces two constraints in problem (\ref{OP4}), i.e., the semi-positive definite constraint (\ref{OP4st1}) and the non-convex rank-one constraint (\ref{OP4st2}). In this case, a penalty function is incorporated into the objective function (\ref{OP4obj}) to enforce the rank-one condition, effectively omitting the constraint (\ref{OP4st2}). Note that both ${{\bf{W}}}$ and ${\bf{V}_m}$ are positive semi-definite matrices with non-negative eigenvalues, as established by constraint (\ref{OP4st1}). Thus, the rank-one constraint (\ref{OP4st2}) can be expressed equivalently as ${\rm{tr}}\left( {{{\bf{W}}}} \right) - {\lambda _{\max }}\left( {{{\bf{W}}}} \right) = 0$ and ${\rm{tr}}\left( {{{\bf{V}}_m}} \right) - {\lambda _{\max }}\left( {{{\bf{V}}_m}} \right) = 0$, where ${\lambda _{\max }}\left(\cdot \right)$ denotes the maximum eigenvalue of a matrix. Further, to smooth the operation of taking the maximum eigenvalue ${\lambda _{\max }}\left(\cdot \right)$, we have the approximate inequalities ${\rm{tr}}\left( {\bf{W}} \right) - {\left( {{\bm{\iota }}_W^\# } \right)^{\rm H}}{\bf{W}} \bm{\iota}_W^\#  \ge {\rm{tr}}\left( {\bf{W}} \right) - {\lambda _{\max }}\left( {\bf{W}} \right) \ge 0$ and ${\rm{tr}}\left( {{\bf{V}}_m^{}} \right) - {\left( {{\bm{\iota }}_{V,m}^\# } \right)^{\rm H}}{\bf{V}}_m^{}{\bm{\iota }}_{V,m}^\#  \ge {\rm{tr}}\left( {{\bf{V}}_m^{}} \right) - {\lambda _{\max }}\left( {{\bf{V}}_m^{}} \right) \ge 0$, where ${{\bm{\iota }}_W^\# }$ and ${{\bm{\iota }}_{V,m}^\# }$ denote unit norm eigenvectors corresponding to the maximum eigenvalues ${\lambda _{\max }}\big( {{\bf{W}}} \big)$ and ${\lambda _{\max }}\left( {{\bf{V}}_{m}} \right)$ in the last iteration, respectively. Eventually, the modified optimization problem with a penalty function is mathematically expressed as problem (\ref{OP5}), presented at the top of the next page.
\begin{figure*}[!t]
\vspace{-13pt}
\begin{subequations}\label{OP5}
\begin{align}
\mathop {\min }\limits_{\scriptstyle{\bf{W}},{{{\bf{V}}}_m},{{\bf{U}}_m},\hfill\atop
\scriptstyle{{\bf{\Omega }}_m},{{\Omega}'_m},{{{\bf{\Omega }}}''_m}\hfill} &\frac{1}{M}\sum\limits_{m = 1}^M {\left( {\rho _m^P{\rm{tr}}\left( {{{\bf{\Omega }}_m}} \right) + \rho _m^T{{\Omega '}_m} + \rho _m^V{\rm{tr}}\left( {{{{\bf{\Omega ''}}}_m}} \right)} \right)}\notag\\
&\ \ \ \ \ \ \ \ \ \ \ \ \ \ \ \ \ \ \ \ \ \ +\rho '\left( {\left( {{\rm{tr}}\left( {\bf{W}} \right) - {{\left( {{\bm{\iota }}_W^\# } \right)}^{\rm H}}{\bf{W }} {\bm{\iota}}_W^\# } \right) + \sum\limits_{m = 1}^M {\left( {{\rm{tr}}\left( {{\bf{V}}_m^{}} \right) - {{\left( {{\bm{\iota }}_{V,m}^\# } \right)}^{\rm H}}{\bf{V}}_m^{}{\bm{\iota }}_{V,m}^\# } \right)} } \right)\label{OP5obj}\\
{\rm{s.t.}}&\ \ (\ref{OP2st1}), (\ref{OP3st1}), (\ref{OP3st2}), (\ref{OP3st3}), (\ref{power_SDR}), (\ref{SAINR_SDR}), (\ref{OP4st1}). \notag
\end{align}
\end{subequations}
\vspace{-11pt}
\hrulefill
\end{figure*}
Herein, $\rho'$ indicates the penalty factor that is increased by a amplification coefficient $\varpi$ in each iteration, significantly impacting the solution accuracy. It is evident that the optimization problem (\ref{OP5}) incorporating the penalty function becomes a standard convex optimization problem when $\mathbf{R}$ is fixed, and thus it can be efficiently solved using readily available convex optimization toolkits.
Ultimately, with the optimal solutions ${\bf{W}}^{*}$ and ${\bf{V}}_m^{*}$ obtained from the iteration process of solving (\ref{OP5}) and updating $\mathbf{R}$, the solution to the original problem (\ref{OP1}) can be calculated through the eigenvalue decomposition (EVD) method, i.e.,
\begin{equation}\label{EVD}
\tilde{{\bf{w}}}^* = \sqrt {{\lambda _{\max }}\left( {{\bf{W}}^*} \right)} {\bm{\iota}}_{W}^*,\  {{\tilde{\bf{v}}_{m}}^*} = \sqrt {{\lambda _{\max }}\left( {{{\bf{V}}_m^*}} \right)} {\bm{\iota}}_{V,m}^*.
\end{equation}
In summary, the proposed penalty function-based beamforming design for integrated navigation and remote sensing in LEO satellite constellations is outlined in Algorithm 1.
\begin{algorithm}
\setstretch{0.9}
\caption{: Beamforming Design for Integrated Navigation and Remote Sensing in LEO Satellite Constellations}
\label{alg1}
\hspace*{0.02in} {\bf Input:} 
$K, M, N_x, N_y, L_k, \varpi, \rho', f', P_k^{\max}, {\eta ^{RS}}, B, T,$ \\
$\text{\ \ \ \ \ \ \ \ } g'_k, \tilde{g}_{k,l_k}, \sigma_s^2, \sigma_m^2, \varrho $.\\
\hspace*{0.02in} {\bf Output:} 
$\tilde{\bf{ w}}$, ${{\tilde{\bf{ v}}}_m}$.
\begin{algorithmic}[1]
\STATE{\textbf{Initialize} iteration index $j=1$, initial feasible points ${{\bf{W}}^{\left( {0} \right)}}$ and ${{\bf{V}}_m^{\left( {0} \right)}}$.}
\REPEAT
\STATE{Update $\mathbf{R}$ according to equation (\ref{R_SDR});}
\STATE{Update ${\bf{W}}^{\#}={\bf{W}}^{\left( j-1 \right)}$ and ${\bf{V}}_m^{\#}={{\bf{V}}_m^{\left( {j-1} \right)}}$;}
\STATE{Obtain ${\bf{W}}^{\left( j \right)}$ and ${{{\bf{V}}}_m}^{\left( j\right)}$ by solving problem (\ref{OP5}) with fixed $\mathbf{R}$;}
\IF{${\bf{W}}^{\left( j \right)}$ and ${{{\bf{V}}}_m}^{\left( j\right)}$ converge}
\IF{$\left|{\rm{tr}}\left( {\bf{W}}^{\left( j \right)} \right) - {\lambda _{\max }}\left( {\bf{W}}^{\left( j \right)} \right)\right| + \sum\limits_{m = 1}^M \left|{\rm{tr}}\left( {{\bf{V}}}^{\left( j \right)}_m \right) - {\lambda _{\max }}\left( {{\bf{V}}}^{\left( j \right)}_m \right)\right|> \varrho$}
\STATE{Update penalty factor $\rho'=\varpi\rho'$};
\ENDIF
\ENDIF
\STATE{Update $j=j+1$;}
\UNTIL Convergence
\STATE{Obtain $\tilde{{\bf{w}}}^*$ and ${{\tilde{\bf{v}}_{m}}^*}$ by the EVD method according to (\ref{EVD}).}
\end{algorithmic}
\end{algorithm}
\vspace{-11pt}

\subsection{Algorithm Analysis}
Herein, we present a detailed analysis of the convergence and computational complexity of the proposed algorithm.

\emph{Convergence Analysis:}
For Algorithm 1, which iteratively solves the convex problem (\ref{OP5}) to obtain a feasible solution, the inequality
$\mathcal{F}\big({{\bf{W}}^{\left( {j} \right)}}, {{\bf{V}}_m^{\left( {j} \right)}}\big) \leq \mathcal{F}\big({{\bf{W}}^{\left( {j-1} \right)}}, {{\bf{V}}_m^{\left( {j-1} \right)}}\big)$  holds at each iteration, where $\mathcal{F}\big({{\bf{W}}^{\left( {j} \right)}}, {{\bf{V}}_m^{\left( {j} \right)}}\big)$ represents the objective value of problem (\ref{OP5}) at the $j$-th iteration. The convergence of Algorithm 1 is guaranteed by the monotone bounded criterion, as the transmit power constraint (\ref{power_SDR}) of LEO satellites and the minimum SAINR required constraint (\ref{SAINR_SDR}) for remote sensing at the central satellite ensure a lower bound on the weighted average PVT error across all UEs \cite{Convergence analyse}. To further validate the convergence behavior of the proposed algorithm, Fig. \ref{Convergence behavior} provides an intuitive illustration under various numbers of LEO satellite collaboration scenarios.

\emph{Complexity Analysis:}
It can be observed that Algorithm 1 is an iterative algorithm, with each iteration involving identical execution steps. Therefore, we focus on analyzing the computational complexity of a single iteration of the proposed algorithm. Specifically, the computational complexity of Algorithm 1 primarily arises from step 5, where the optimal solutions for ${{\bf{W}}}$ and ${{\bf{V}}_m}$ are obtained by solving problem (\ref{OP5}). Since the convex problem (\ref{OP5}) includes only LMI constraints, it can be efficiently solved using a standard interior-point method (IPM). Consequently, the worst-case runtime of the IPM can be employed to characterize the computational complexity of the proposed algorithm. In particular, problem (\ref{OP5}) has $M$ LMI constraints of dimension $K+1$, $2M$ LMI constraints of dimension $K+3$, $M$ LMI constraints of dimension $2K$, $M+K+2$ LMI constraints of dimension $NK$, and $\varsigma=N^2K^2+N^2K^2M+K^2+19$ optimization variables. As a result, for the solution with a given precision $\zeta  > 0$, the worst-case complexity for solving the problem (\ref{OP5}) for per iteration is denoted as $\sqrt {N{K^2} + NKM + 5KM + 2NK + 7M} \Xi \ln \left( {1/\zeta } \right)$, where $\Xi=\varsigma ( ( 6{K^3}M + 18{K^2}M + 65KM + 55M + {N^3}{K^3}M + {N^3}{K^4}+ 2{N^3}{K^3} ) + \varsigma ( {7{K^2}M + 14KM + 19M + {N^2}{K^2}M + {N^2}{K^3} + 2{N^2}{K^2}} ) )$ with decision variable $\varsigma=\mathcal{O} (N^2K^2M)$ \cite{complexity analysis}. This analysis confirms the polynomial-time nature of the algorithm and its theoretical tractability in practical application scenarios, with respect to the key system parameters $N$, $K$, and $M$.

\section{Simulation Results}
This section describes the parameter settings used in the numerical simulations and presents the results to validate the effectiveness of the proposed algorithm. Without loss of generality, we adopt a classical Walker Delta constellation similar to Starlink program, with detailed parameters listed in Table II \cite{Walker Delta1}. Specifically, a service group of satellites from the LEO satellite constellation is utilized for the simulations. It is assumed that the positions of navigation UEs and the remote sensing area are randomly distributed within the coverage area corresponding to central satellite elevation angles ranging from $50^\circ$ to $90^\circ$. Furthermore, the ambiguity areas are assumed to be randomly distributed within a 10 km radius centered on the remote sensing area. Unless otherwise specified, the simulation parameters are configured as detailed in Table III.

\begin{table}[ht]
\vspace{-7pt}
\small
\centering
\caption{Parameters Of LEO Satellite Constellation }\label{Simulation}
\vspace{-6pt}
\begin{tabular}{|c|c|}
\hline
Parameter & Value  \\ \hline
Orbital altitude & $h^{*}=550$ km \\\hline
Number of orbital planes & $P^{*}=72$ \\\hline
Total satellites & $N^{*}=1296$ \\\hline
Orbital inclination & $i^{*}=53^{\circ}$ \\\hline
Phase factor & $F^{*}=45$ \\\hline
\end{tabular}
\vspace{-5pt}
\end{table}

\begin{table}[ht]
\vspace{-7pt}
\small
\centering
\caption{Simulation Parameters Configuration}\label{Simulation}
\vspace{-6pt}
\begin{tabular}{|c|c|}
\hline
Parameter & Value \\ \hline
Number of UPA antennas & $N=N_x\times N_y=4\times 4$ \\\hline
Number of LEO satellites in a group & $K=5$ \\\hline
Number of UEs covered & $M=10$ \\\hline
Speed of light & $c=3\times 10^{8}$ m/s \\\hline
Signal frequency & $f'=35$ GHz \\\hline
Boltzmann constant & $\kappa=1.38 \times 10^{-23}$ J/m \\\hline
Channel bandwidth & $B=20$ MHz \\\hline
Receive gain & $G_m=55$ dBi \\\hline
Noise temperature & $T=100$ K \\\hline
Rain attenuation mean & $\mu _r=-2.6$ dB \\\hline
Rain attenuation variance & $\sigma _r^2=1.63$ dB \\\hline
Maximum antenna gain & ${b_{k}^{\max}}=16$ dBi \\\hline
3-dB angle & $\varepsilon_k^{3dB}=0.4^{\circ}$  \\\hline
Noise power & $\sigma_s ^2=-110$ dBm \\\hline
Antenna spacing to signal wavelength & $d=\frac{1}{2}\lambda$ \\\hline
Maximum transmit power & $P_k^{\max }=30$ dBm \\\hline
PVT error weights & $\rho_m ^P=1,$\\ &$\rho_m ^V=10$, $\rho_m ^T=10^9$ \\\hline
Minimum required SAINR & $\eta ^{RS}=10$ dB \\\hline
Number of ambiguity areas & $L_k=5$ \\\hline
Initial penalty factor & $\rho'=10$ \\\hline
Amplification coefficient & $\varpi=1.5$ \\\hline
Penalty accuracy & $\varrho=10^{-4}$ \\\hline
\end{tabular}
\end{table}

\begin{figure}[!t]
\centering
\includegraphics [width=0.4\textwidth] {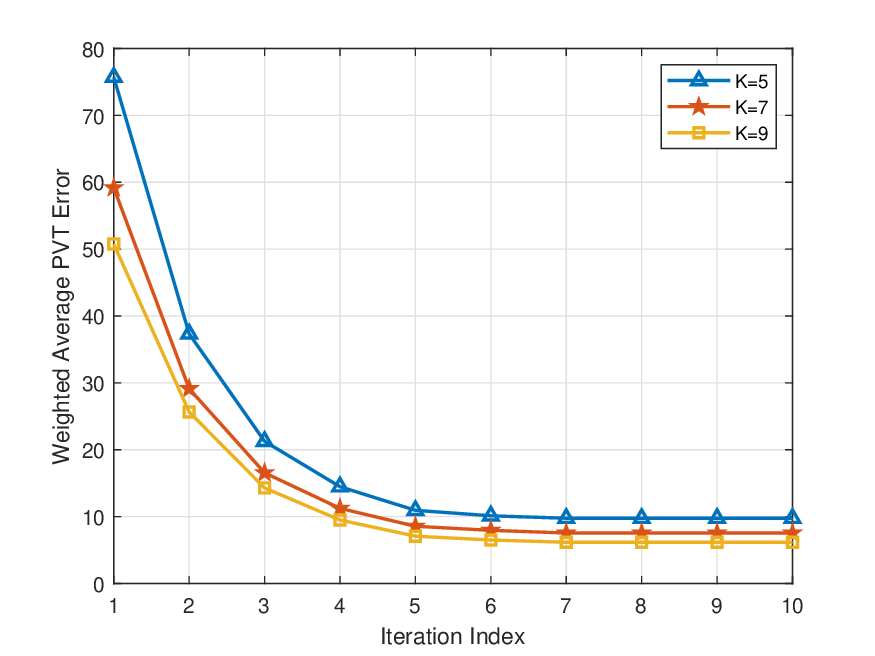}
\caption {Convergence behavior.}
\label{Convergence behavior}
\end{figure}
Firstly, Fig. \ref{Convergence behavior} demonstrates the convergence behavior of Algorithm 1 under various numbers of dual-function LEO satellites in collaboration. The results show that the weighted average PVT error decreases progressively over iterations and stabilizes within 10 iterations. This highlights that the computational complexity of the proposed algorithm is feasible for practical implementation, enabling the integration of high-precision navigation and high-quality remote sensing in LEO satellite constellations. Furthermore, the navigation performance improves as the number of collaborative satellites in the service group increases.

\begin{figure}[!t]
\centering
\includegraphics [width=0.4\textwidth] {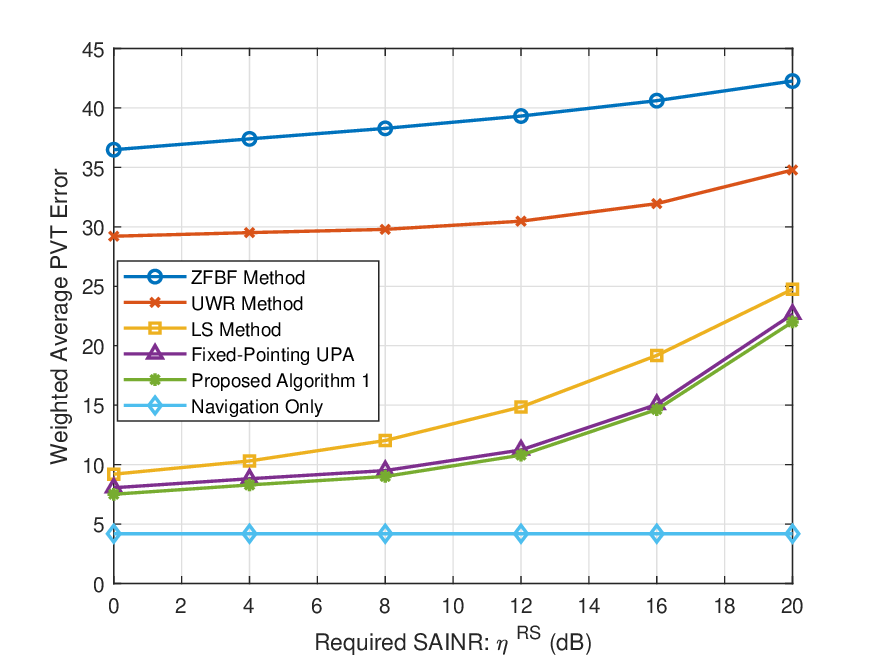}
\caption {Weighted average PVT error versus required SAINR for different integration designs.}
\label{compare}
\end{figure}
Secondly, Fig. \ref{compare} presents the performance improvements achieved by the proposed Algorithm 1 in comparison with other integration designs. Specifically, the ``ZFBF Method" utilizes the zero-forcing beamforming approach for the design of navigation and remote sensing beamforming \cite{ZFBF Method}. The ``UWR method" applies uniformly weighted reception (UWR) in the receive beamforming design at the central satellite \cite{UWR}. The ``LS method" assigns equal observational weights to all satellites during pseudo-range measurements. The ``Fixed-Pointing UPA" represents a configuration where all satellites in the service group have UPAs fixed to point in the same direction \cite{CRB ref2}. Lastly, the ``Navigation Only" approach evaluates the upper bound of navigation performance by excluding the SAINR requirement constraint (\ref{OP1st2}) when solving problem (\ref{OP1}). The proposed Algorithm 1 demonstrates superior performance overall. While its performance is comparable to that of the ``Fixed-Pointing UPA", the ``Fixed-Pointing UPA" design is overly idealized and fails to meet the requirements for global coverage. In contrast, the geocentric pointing UPA is more suitable for dynamic global coverage scenarios, ensuring continuous navigation and remote sensing capabilities. Additionally, the gap between the ``Proposed Algorithm 1" and ``Navigation Only" lines highlights the loss in PVT accuracy caused by accommodating remote sensing services. Therefore, the proposed Algorithm 1 offers a practical and efficient solution for integrated navigation and remote sensing in dynamic global coverage scenarios.

\begin{figure}[!t]
\centering
\includegraphics [width=0.4\textwidth] {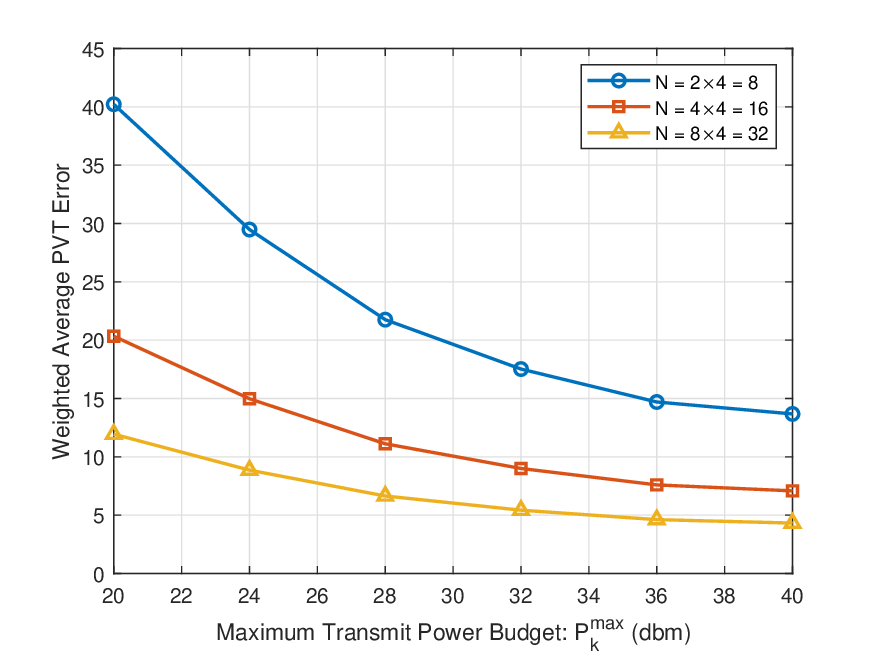}
\caption {Weighted average PVT error versus the maximum transmit power budget for different sizes of UPA.}
\label{UPA sizes}
\end{figure}
Next, Fig. \ref{UPA sizes} explores how the system performance is influenced by various UPA sizes on LEO satellites and the maximum satellite transmit power budgets. The results show that the weighted average PVT error reduces as the maximum transmit power budget $P_k^{\max}$ increases. This improvement arises from the fact that dual-function signals, which suffer significant attenuation in satellite-terrestrial channels, achieve higher quality at the receiver when transmit power is increased, thereby enhancing both navigation and remote sensing accuracy. Furthermore, the weighted average PVT error decreases with the scale of the UPA on LEO satellites. Larger UPAs boost the efficiency of navigation signal transmission and offer additional spatial degrees of freedom for remote sensing. Therefore, deploying UPAs of an optimal size on LEO satellites is crucial to balancing construction costs and overall system performance in practical applications.

\begin{figure}[!t]
\centering
\includegraphics [width=0.4\textwidth] {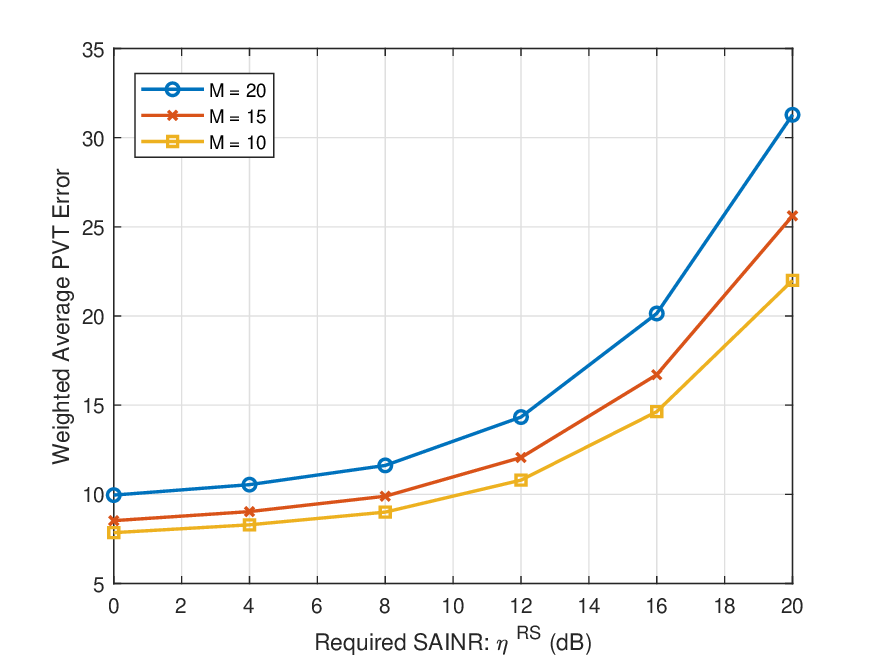}
\caption {Weighted average PVT error versus required SAINR for different numbers of UE.}
\label{UE number}
\end{figure}

\begin{figure}[!t]
\centering
\includegraphics [width=0.4\textwidth] {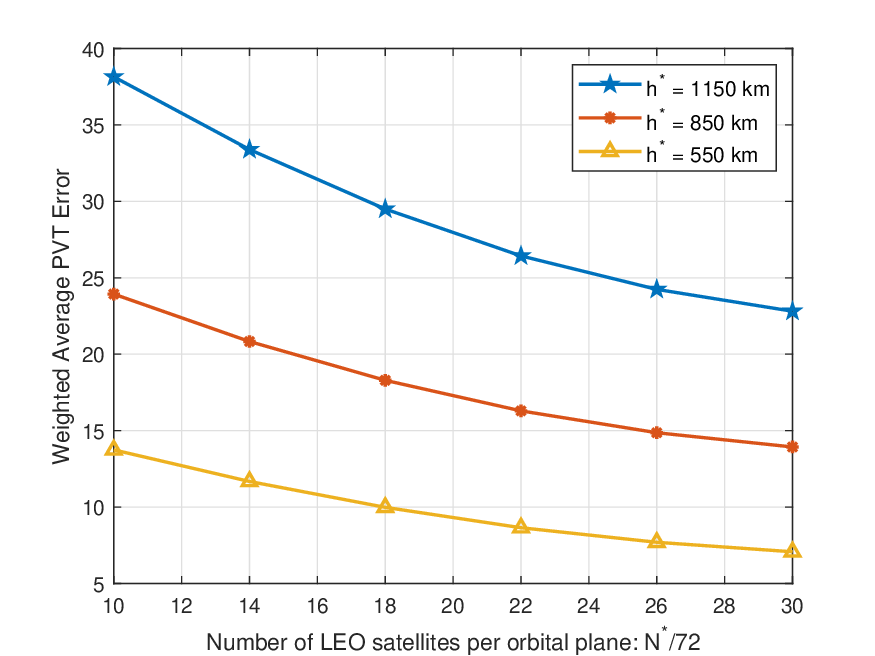}
\caption {Weighted average PVT error versus the number of LEO satellites per orbital plane for different orbital altitudes of satellite constellations.}
\label{satellite h}
\end{figure}

Fig. \ref{UE number} depicts the relationship between the weighted average PVT error and the required SAINR for different numbers of navigation UEs. Specifically, the number of UEs has a significant impact on system performance, with a higher number of UEs leading to a more pronounced decline in PVT accuracy. This degradation is primarily due to the additional navigation interference introduced as the number of UEs increases. Moreover, a clear trade-off between navigation and remote sensing performance is observed, as the weighted average PVT error increases significantly with higher SAINR requirements. Thus, it is crucial to balance navigation and remote sensing performance to meet specific operational requirements.

Finally, Fig. \ref{satellite h} illustrates the relationship between the weighted average PVT error and the number of LEO satellites per orbital plane for satellite constellations with various orbital altitudes. The results show that, with a fixed number of orbital planes, increasing the number of satellites per plane leads to a gradual reduction in the weighted average PVT error. This improvement is attributed to the denser deployment of LEO satellites, which significantly reduces the distances between satellites and navigation UEs, as well as between satellites and remote sensing area, thereby improving channel conditions for both navigation and remote sensing. Moreover, lowering the orbital altitudes further enhances PVT accuracy by reducing signal attenuation and transmission delays, as satellites operate closer to the earth's surface. In recent years, satellite constellations have increasingly adopted lower orbital altitudes and larger scales to achieve superior system performance. While these trends offer clear advantages in terms of navigation and remote sensing accuracy, they also present significant challenges. Lower orbital altitudes result in higher atmospheric drag, which shortens satellite lifespans, while larger constellations increase system complexity and deployment costs. Therefore, it is essential to design LEO satellite constellations, considering service requirements and the need for continuous coverage, to strike an optimal balance between performance, cost, and sustainability.

\section{Conclusion}
This paper proposed an integrated framework for navigation and remote sensing in LEO satellite constellations, addressing the growing demand for multi-functional, global coverage satellite systems. A unified signal frame structure was designed to support efficient synchronization between navigation and remote sensing tasks, while enabling seamless sharing of signals within the same spectrum resources. To enhance system performance, a joint beamforming design was introduced to balance the performance metrics, i.e., PVT error for navigation and SAINR for remote sensing. Given the complexity of the non-convex optimization problem, a penalty function-based iterative algorithm was developed to obtain a feasible solution, offering a practical implementation approach. Finally, the theoretical analysis and simulation results validated the feasibility of the proposed integrated navigation and remote sensing framework and beamforming deign algorithm, demonstrating their clear advantages over traditional schemes.

\end{document}